\pdfoutput=1
\documentclass[a4paper,12pt,english]{article}

\usepackage[pdftex]{graphicx}
\usepackage{graphicx}
\usepackage{colordvi}
\usepackage{fancybox}
\usepackage{cancel}
\usepackage{amsmath}
\usepackage{slashed}
\usepackage{amssymb}
\usepackage{caption}
\usepackage{appendix}
\usepackage{hyperref} 
\usepackage{multirow}
\hypersetup{colorlinks=true, linkcolor=black, urlcolor=blue}
\usepackage{float}
\usepackage{cite}
\usepackage{mathtools}
\usepackage{bbm}

\usepackage{wasysym}

\usepackage{color}
\definecolor{gray}{rgb}{0.5,0.5,0.5}

\def\det{\mbox{det}\,}

\newcommand\lsim{\mathrel{\rlap{\lower4pt\hbox{\hskip1pt$\sim$}}
    \raise1pt\hbox{$<$}}}
\newcommand\gsim{\mathrel{\rlap{\lower4pt\hbox{\hskip1pt$\sim$}}
    \raise1pt\hbox{$>$}}}

\newcommand{\beq}{\begin{equation}}
\newcommand{\eeq}{\end{equation}}
\newcommand{\bea}{\begin{eqnarray}}
\newcommand{\eea}{\end{eqnarray}}
\newcommand{\bem}{\begin{pmatrix}}
\newcommand{\eem}{\end{pmatrix}}
\newcommand{\noi}{\noindent}
\newcommand{\non}{\nonumber}

\newcommand{\bet}{\begin{itemize}}
\newcommand{\eet}{\end{itemize}}
\newcommand{\ben}{\begin{enumerate}}
\newcommand{\een}{\end{enumerate}}

\begin{document}

\numberwithin{equation}{section}

\begin{flushright}
\end{flushright}

\bigskip

\begin{center}

{\Large\bf The free and safe fate of symmetry non-restoration}

\vspace{1cm}

\centerline{Borut Bajc$^{a,}$\footnote{borut.bajc@ijs.si}, Adri\'an Lugo$^{b,}$\footnote{lugo@fisica.unlp.edu.ar}
 and Francesco Sannino$^{c,d}$\footnote{sannino@cp3.sdu.dk}}

\vspace{0.5cm}
\centerline{$^{a}$ {\it\small J.\ Stefan Institute, 1000 Ljubljana, Slovenia}}
\centerline{$^{b}$ {\it\small Instituto de F\'isica de La Plata-CONICET, and Departamento de F\'isica,}}
\centerline{{\it\small Facultad de Ciencias Exactas, Universidad Nacional de La Plata, Argentina}}
\centerline{$^{c}$ {\it\small CP$^3$-Origins \& the Danish IAS, University of Southern Denmark,  Denmark}}
\centerline{$^{d}$ {\it \small Dipartimento di Fisica, E. Pancini, Univ. di Napoli, Federico II and INFN sezione di Napoli.}}
\centerline{\it \small  Complesso Universitario di Monte S. Angelo Edificio 6, via Cintia, 80126 Napoli, Italy}
\end{center}

\bigskip

\begin{abstract}
We investigate the high temperature fate of four dimensional gauge-Yukawa theories  featuring short distance conformality of either interacting or non-interacting nature. The latter is known as complete asymptotic freedom and, as templates, we consider non-abelian gauge theories featuring either two singlet scalars coupled to gauged fermions via Yukawa interactions or two gauged scalars with(out) fermions.  For theories with interacting fixed points at short distance, known as asymptotically safe, we consider two calculable examples.
 Exploring the landscape of safe and free theories above we discover  a class of complete asymptotically free theories for which symmetry breaks at arbitrary high temperatures. In its minimal form this class is constituted by a theory  with two fundamental gauged scalars each gauged under an independent group.   
\end{abstract}

\clearpage

\tableofcontents

\section{Introduction}

The phenomenon of symmetry non-restoration (for a review see for example 
\cite{Senjanovic:1998xc,Bajc:1999cn}) has been first noticed by Weinberg \cite{Weinberg:1974hy}
and then studied in detail by Mohapatra and Senjanovi\'c \cite{Mohapatra:1979qt,Mohapatra:1979vr} who 
were the first ones to successfully apply the mechanism to phenomenology. Since then it has been employed in cosmology to address various issues like the monopole  \cite{Langacker:1980kd,Salomonson:1984rh,Dvali:1995cj}, the domain wall \cite{Dvali:1995cc} and false vacuum problems. The phenomenon has also been invoked for other phenomena including baryogenesis \cite{Mohapatra:1979zc,Kuzmin:1981bc,Kuzmin:1981ip,Kuzmin:1982hy,Dodelson:1989ii,Dodelson:1991iv} and inflation \cite{Lee:1995fb}.

Symmetry non-restoration at high energy can occur also due to the concomitance of other mechanisms such as the presence of large charges that can induce either Bose-Einstein condensation or  superconductivity.  This mechanism has been used in the literature \cite{Linde:1976kh,Haber:1981ts,Benson:1991nj,Riotto:1997tf,Liu:1993am,Bajc:1997ky,Bajc:1998rd,Bajc:1999he}. For example a large charge can still be realistically related to the yet to be experimentally determined neutrino lepton number 
\cite{Kang:1991xa,Kinney:1999pd,Lesgourgues:1999wu,Barenboim:2016shh,Barenboim:2016lxv,Barenboim:2017dfq}. 

Symmetry non-restoration at high temperature cannot occur in supersymmetry 
\cite{Haber:1982nb,Mangano:1984dq,Bajc:1996kj} unless we have flat directions \cite{Dvali:1998ct,Bajc:1998jr} and/or nonzero fixed charge. 
 
For non supersymmetric quantum field theories symmetry non-restoration has been tested via different methods in \cite{Bimonte:1995sc,Orloff:1996yn,Roos:1995vm,Pietroni:1996zj,AmelinoCamelia:1996sd,Pinto:1999pg} for global symmetries and non-restoration for local symmetries have been  investigated in \cite{Bimonte:1995xs}. The results seem to support the existence of symmetry non-restoration although these claims have been challenged in \cite{Fujimoto:1984hr,Klimenko:1988ng,Grabowski:1990qc,Gavela:1998ux,Bimonte:1998he}. 

Analyses including generalisation to different space-time dimensions including $\epsilon$ dimensions away from four are summarised in Ref.~\cite{Hong:2000rk,Chai:2020zgq,Buchel:2020thm,Buchel:2020jfs}. More precisely: symmetry non-restoration at high temperature is possible also in lower 
\cite{Hong:2000rk,Buchel:2020thm,Buchel:2020jfs} and non-integer dimensions \cite{Chai:2020zgq}.

A common feature of all the theories studied so far for symmetry non-restoration at high temperature is that these can be viewed as effective theories without a well defined ultraviolet completion. This fact implies that the arbitrary large temperature limit cannot be taken. 

In this work we go one step beyond  with respect to what has been done so far by analysing Weinberg's symmetry non-restoration hypothesis within models that are well defined at short distance. These are, according to Wilson  \cite{Wilson:1971bg,Wilson:1971dh} and Weinberg \cite{Weinberg} classification of well defined theories,  of either asymptotically free or safe nature.  Within these theories it is consistent to consider the infinite temperature limit. 
It is worth recalling that for these theories short scale conformality guarantees the existence of a well defined theory at high energy making them UV complete. 
 Asymptotic safety for gauge-Yukawa theories was discovered in \cite{Litim:2014uca} with the corrections to the quantum potential presented in \cite{Litim:2015iea}.  Interestingly, once asymptotic freedom is lost in the gauge-fermion sector, within perturbation theory, the fundamentality of the theory can only be reinstated via Yukawa interactions. This implies that elementary scalars are needed, for the first time, to tame the high energy behaviour of the theory.  The discovery of asymptotic safe quantum field theories \cite{Litim:2014uca} has led to an ongoing number of theoretical \cite{Intriligator:2015xxa,Bajc:2016efj,Abel:2017ujy,Bajc:2017xwx,Orlando:2019hte} and phenomenological investigations \cite{Mann:2017wzh,Pelaggi:2017abg,Abel:2017rwl,Bond:2017wut,Cacciapaglia:2018avr,Cacciapaglia:2019vcb,Sannino:2019sch,Molinaro:2018kjz},
  including the recent discovery of safe non-supersymmetric grand unified theories of \cite{Fabbrichesi:2020svm} which naturally integrates and complements the supersymmetric story of \cite{Bajc:2016efj}.

  \vskip .3cm
For the issue of symmetry non-restoration at arbitrary high temperatures we consider, at first, the landscape of complete asymptotically free non-abelian gauge theories that feature either two singlet scalars coupled to gauged fermions via Yukawa interactions or two gauged scalars {without  Yukawa interaction}.  

The first model we encounter of complete asymptotically free theories for which symmetry breaks at arbitrary high temperatures is  constituted by  two gauged scalars transforming according to the fundamental representation of two distinct gauge groups with fermions also transforming in the fundamental representation and without Yukawa interactions.  
 
To investigate the high-temperature fate of global symmetries for asymptotically safe theories we consider the Litim-Sannino model of  \cite{Litim:2014uca} and one of its variations that has been used for perturbative safe extensions of the standard model \cite{Abel:2018fls}.   We show that for these examples  the safe quantum global symmetries   are restored at  high temperatures.

\section{Complete asymptotically free theories at high temperature}

Before embarking in our main quest, which is to investigate the symmetry (non)restoration phenomenon for complete asymptotically free quantum field theories we briefly summarise Weinberg's (non-free) model mechanics.  In its most minimal form  the model features two scalars with the following 
quartic potential

\beq
\label{weinbergV}
V=\frac{\lambda_1}{4}\phi_1^4+\frac{\lambda_2}{4}\phi_2^4-\frac{\lambda}{2}\phi_1^2\phi_2^2
\eeq

\noi
with discrete $Z_2\times Z_2$ symmetry $\phi_1\to-\phi_1$ and $\phi_2\to-\phi_2$. For 

\beq
\label{boundedness}
\lambda_{1,2}>0\;\;\;,\;\;\;\lambda^2<\lambda_1\lambda_2
\eeq

\noi
the model is bounded from below. At high temperature the following correction arises: 
\cite{Weinberg:1974hy}:

\beq
\label{deltaVT}
\Delta V_T=\frac{T^2}{24}\left((3\lambda_1-\lambda)\phi_1^2+(3\lambda_2-\lambda)\phi_2^2\right) \ .
\eeq
With $\lambda>3\lambda_2$ (but with $\lambda_1$ satisfying  (\ref{boundedness})) the field $\phi_2$ 
acquires a negative thermal mass squared at  high temperature which yields a non-zero vev $\langle\phi_2\rangle\ne0$. Therefore in this case the second $Z_2$  breaks at sufficiently high temperatures. This theory is, however, not UV complete since the scalar couplings increase with the energy. Assuming a physical cutoff, for temperatures below this cutoff one therefore observes the phenomenon of symmetry non-restoration.   

Because the theory is limited by a physical cutoff we cannot ask the relevant question of whether the symmetry remains broken at arbitrary high temperatures. 
This is exactly what our work wishes to achieve, i.e. what is the ultimate fate of the symmetry in a truly UV complete theory (up to gravity) at arbitrary large temperatures.  

Here we analyse complete asymptotically free theories that are natural UV completions of the Weinberg's model.  These require the presence of gauge fields and the gauge sector to be asymptotically free given that it is this sector  the one responsible to drive the Yukawa and scalar couplings to be asymptotically free as well.

We  divide our theories in whether they feature gauge singlets or gauged scalars.

\subsection{Symmetry restoration with singlet scalars}

To start with we consider an $SU(N_c)$ gauge group with $N_f=N_{f_1}+N_{f_2}$ Dirac fermions in the fundamental representation coupled  
to the scalars $\phi_{1,2}$ via the following $Z_2\times Z_2$ symmetry ($\phi_k\to-\phi_k$, $\psi_k\to i\gamma_5\psi_k$) preserving Yukawa terms:
\beq
{\cal L}_Y=\phi_1\sum_{i=1}^{N_{f_1}}y_{1i}\overline{\psi}_{1i}\psi_{1i}
+\phi_2\sum_{i=1}^{N_{f_2}}y_{2i}\overline{\psi}_{2i}\psi_{2i} \ .
\eeq
 
Because we are searching for asymptotically free solutions we must have that  $\alpha_g\propto1/t$ for large $t=\log{(\mu/\mu_0)}$ with $\mu$ the renormalisation scale and $\mu_0$ a reference scale.  { Complete asymptotic freedom requires that all  couplings must vanish at infinity at least as fast as  $\alpha_g$ and therefore their scaling must be proportional to $1/t^a$ with $a\geq 1$.  Additionally the requirement of a negative thermal mass given in \eqref{m2T}, necessary for symmetry breaking, implies that at least some scalar quartic couplings cannot decrease faster than the gauge coupling, i.e. they must approach zero as $1/t^b$ with $b\leq1$. }
 Therefore, for the purpose of our work, it is sufficient to investigate the fixed flow solution according to which all couplings vanish at infinity as $1/t$ \cite{Giudice:2014tma}. This observation greatly simplifies the following analyses by transforming a set of non-linear and coupled first order ordinary differential equations into a system of 
non-linear and coupled {\it polynomial}  equations. In practice, by defining  ($g$ is the gauge coupling)
\beq
\alpha_g=\frac{g^2}{(4\pi)^2}\;\;\;,\;\;\;
\alpha_{y_i}=\frac{y_i^2}{(4\pi)^2}\;\;\;,\;\;\;
\alpha_{\lambda_i}=\frac{\lambda_i}{(4\pi)^2}\;\;\;,\;\;\;
\alpha_{\lambda}=\frac{\lambda}{(4\pi)^2}
\eeq

\noi
($i=1,2$) we will search for solutions of the asymptotic form

\beq
\alpha_a=\frac{\tilde\alpha_a}{t}\;\;\;,\;\;\;a=g,y_1,y_2,\lambda_1, \lambda_2,\lambda,\ldots 
\label{tildecouplings}
\eeq
with constant $\tilde{\alpha}_a$. 

We are now ready to investigate the first relevant examples with singlets scalars and then we will generalise the results to a wider class of theories.

\subsubsection{ SU($N_c$)   with two singlet scalars and fundamental fermions  }

In this model, described in detail in Appendix \ref{SUNcsinglets}, we consider  two singlet scalars coupled through Yukawa interactions 
to $N_{f_1}$ ($N_{f_2}$) Dirac fermions in the fundamental representation of $SU(N_c)$. We further allow for $N_{f_0}$  
  Dirac fermions in the fundamental representation of the gauge group that are inert with respect to the scalars, i.e. do not possess Yukawa couplings. 

\vskip .3cm
We  now provide an elegant proof that at high temperature this theory, if complete asymptotically free, cannot break any symmetry. 
Let us start with the thermal masses for the scalars that at one loop read (\ref{mTNc}): 
\beq
m_i^2(T)=(4\pi)^2\frac{T^2}{12\log{T}}\left(3\tilde\alpha_{\lambda_i}-\tilde\alpha_{\lambda}+2N_cN_{f_i}\tilde\alpha_{y_i}\right) \ ,
\eeq
written in terms of  \eqref{tildecouplings} couplings. It is sufficient to consider one of the two scalar masses to be negative. Here we choose to be $m^2_1$ that requires 
\beq
\label{pozitivenq}
\tilde\alpha_{\lambda}-2N_cN_{f_1}\tilde\alpha_{y_1}>3\tilde\alpha_{\lambda_1}>0 \ .
\eeq 
Under the assumption that there is a complete asymptotically free solution we have (\ref{eqlambda1}) 
\beq
\label{eqlambdaq1}
-\tilde\alpha_{\lambda_1}=18\tilde\alpha_{\lambda_1}^2+2\tilde\alpha_{\lambda}^2
-8N_cN_{f_1}\tilde\alpha_{y_1}^2+8N_cN_{f_1}\tilde\alpha_{y_1}\tilde\alpha_{\lambda_1} \ , 
\eeq
for the relevant scalar coupling as function of the other couplings. The general form of the RGE equations can be found in the Appendix 
\ref{SUNcsinglets}. 

Rewriting  (\ref{eqlambdaq1}) as
\beq
\label{enacbaq}
2\left(\tilde\alpha_{\lambda}^2-4N_cN_{f_1}\tilde\alpha_{y_1}^2\right)+
\tilde\alpha_{\lambda_1}+18\tilde\alpha_{\lambda_1}^2
+8N_cN_{f_1}\tilde\alpha_{y_1}\tilde\alpha_{\lambda_1}=0 \ ,
\eeq
we notice that every term except the first one is positive. This means that  to satisfy this equation, the 
first term  must be negative for the fixed flow solution to be possible.  However, since the first term can be rewritten as
\beq
\tilde\alpha_{\lambda}^2-4N_cN_{f_1}\tilde\alpha_{y_1}^2=\left(\tilde\alpha_{\lambda}-2N_cN_{f_1}\tilde\alpha_{y_1}\right)
\left(\tilde\alpha_{\lambda}+2N_cN_{f_1}\tilde\alpha_{y_1}\right)+4N_cN_{f_1}\left(N_cN_{f_1}-1\right)\tilde\alpha_{y_1}^2 \ ,
\eeq
the simultaneous requirement of the presence of a negative mass squared term   implies that also the first term is positive due to (\ref{pozitivenq}). 

We have therefore shown that \eqref{enacbaq}  cannot 
have a solution and that the symmetry must be restored for this model at high temperature once complete asymptotic freedom is enforced.

\subsubsection{More general result for singlet scalars}
Let us consider the  more general scalar potential 
\beq
V=\frac{\lambda}{4}\left(\phi^T\phi\right)^2-\frac{1}{2}\left(\phi^T\phi\right)\eta_{ij}\chi^i\chi^j+V(\chi)
\eeq

\noi
where $\phi$ is a real vector with  $d_\phi$ components. The global symmetry at the potential level over $\phi$ is $O(d_\phi)$.  Under this group $\phi$ transforms with a $d_\phi \times d_\phi$ orthogonal matrix $O$ of $O(d_\phi)$ as:  
\beq
\label{O}
\phi'=O\phi \ .
\eeq

\noi
We further  consider an arbitrary gauge group with Weyl fermions transforming according to an arbitrary gauge representation compatible with asymptotic freedom \cite{Dietrich:2006cm}.  The Yukawa terms written directly in terms of the Weyl fermions read: 
\beq
{\cal L}_{Yukawa}=\frac{1}{2}\phi^a\psi_iY^a_{ij}\psi_j+h.c.+{\cal L}_{Yukawa}(\chi,\psi') \ .
\label{Yukawa-gen}
\eeq

\noi
Under the assumption that 
\beq
\label{yukinv}
\psi'=U\psi\quad,\quad  \widetilde{O}^a_bU_{ki}Y^b_{kl}U_{lj}=Y^a_{ij}
\eeq
with $\widetilde{O}$ a rotation matrix that is part of a subgroup of $O(d_\phi)$ and $U$ a unitary transformation with $i,j,k = 1, \cdots, N_f$ with $N_f$ the number of Weyl matter fields, the Yukawa terms preserves the resulting subgroup of $O(d_\phi)$. 
The information on which fermions couple to $\phi$ is clearly hidden in the Yukawa matrix. The last unspecified Yukawa terms in \eqref{Yukawa-gen} contain interactions of the $\chi$ scalar fields with the Weyl fermions $\psi^\prime$ that are not coupled to $\phi$.  
We now show that the thermal mass of $\phi$ cannot be negative at high temperatures when the theory is required to be asymptotically free also in all couplings.

Let us consider the thermal mass  
\beq
\label{mass}
m_\phi^2(T)=(4\pi)^2\frac{T^2}{12\log{T}}\left((d_\phi+2)\tilde\lambda-\tilde\eta_{kk}+Tr\left(\tilde Y_1^\dagger\tilde Y_1\right)\right) \ ,
\eeq

\noi
where we used

\beq
Tr\left(Y^{a\dagger}Y^b\right)=\delta^{ab}Tr\left(Y_1^\dagger Y_1\right) \ .
\eeq

\noi
and defined as usual

\beq
\lambda=(4\pi)^2\frac{\tilde\lambda}{t}\quad,\quad\eta_{ij}=(4\pi)^2\frac{\tilde\eta_{ij}}{t}
\quad,\quad Y^a=4\pi\frac{\tilde Y^a}{t^{1/2}}
\eeq

\noi
with $\tilde\lambda$, $\tilde\eta_{ij}$, $\tilde Y^a$ constants.

In order not to restore the symmetry carried by the potential term and the Yukawa relative to $\phi$ the thermal mass (\ref{mass}) must be negative. This implies: 

\beq
\label{negativemass}
\tilde\eta_{kk}-Tr\left(\tilde Y_1^\dagger \tilde Y_1\right)>\left(d_\phi+2\right)\tilde\lambda>0 \ .
\eeq

Let us now compute the RGE for $\tilde\lambda$ relative to achieving the fixed flow solution:

\beq
2\left(\tilde\eta_{ij}\tilde\eta_{ij}-2Tr\left(\tilde Y_1^\dagger \tilde Y_1\tilde Y_1^\dagger \tilde Y_1\right)\right)+
2(d_\phi+8)\tilde\lambda^2+\tilde\lambda+4\tilde\lambda Tr\left(\tilde Y_1^\dagger \tilde Y_1\right)=0 \ ,
\eeq 

\noi
where we used 

\beq
Tr\left(Y^{a\dagger}Y^bY^{c\dagger}Y^d\right)=A\delta_{ab}\delta_{cd}+B\delta_{ac}\delta_{bd}+C\delta_{ad}\delta_{bc}\ ,
\eeq

\noi
which follows from the symmetry properties of the Yukawa matrices (\ref{yukinv}).

\vskip .3cm
To obtain a solution, the first term must be negative (all the others are positive). However, we have

\bea
\label{eta}
\tilde\eta_{ij}\tilde\eta_{ij}-2Tr\left(\tilde Y_1^\dagger \tilde Y_1\tilde Y_1^\dagger \tilde Y_1\right)&=&
\left(\tilde\eta_{kk}-Tr\left(\tilde Y_1^\dagger \tilde Y_1\right)\right)\left(\tilde\eta_{kk}+Tr\left(\tilde Y_1^\dagger \tilde Y_1\right)\right)\\
&+&2\sum_{i<j}\tilde\eta_{ij}\tilde\eta_{ij}+\left(\left(Tr\left(\tilde Y_1^\dagger \tilde Y_1\right)\right)^2-2Tr\left(\tilde Y_1^\dagger \tilde Y_1\right)^2\right)\non \ .
\eea

The first term on the right-hand-side is positive due to the assumption of the occurrence of a negative thermal mass squared (\ref{negativemass}), therefore the only  
possible negative term could be the last one. Since the above traces are invariant under unitary rotations of the Hermitian matrix 
$\tilde Y_1^\dagger \tilde Y_1$, we are free to consider the basis with diagonal 

\beq
\left(\tilde Y_1^\dagger \tilde Y_1\right)_{ij}=\tilde y_{1i}^2\delta_{ij}
\eeq

\noi
so that (\ref{eta}) becomes

\bea
\left(Tr\left(\tilde Y_1^\dagger\tilde Y_1\right)\right)^2-2Tr\left(\tilde Y_1^\dagger \tilde Y_1\right)^2&=&
\sum_{\mu}dim\left(R_\mu^1\right)\left(dim\left(R_\mu^1\right)-2\right)\tilde y_{1\mu}^4\non\\
&+&
2\sum_{\mu<\mu'}dim\left(R_\mu^1\right)dim\left(R_{\mu'}^1\right)\tilde y_{1\mu}^2\tilde y_{1\mu'}^2
\eea

\noi
with $\mu$ and $\mu'$ running over the fermion representations. For non gauge singlet fermions we have $dim\left(R_{\mu'}\right) \geq 2$ and therefore the right hand side is positive. For gauge singlet fermions the only solution compatible with a UV well defined theory is the one for which the Yukawa coupling vanishes identically and therefore the previous equation does not apply. 

 Therefore there is no solution to the RGE for $\tilde\lambda$. Or, in other words, 
if a fixed flow solution exists, it cannot have a negative thermal mass. 
The previous example with a $Z_2$ symmetry is included here by assuming the original symmetry to be simply a $Z_2$
for $d_\phi=1$.

\subsection{Exploring symmetry non-restoration with gauged scalars}
So far we have shown that a great deal of gauge theories with scalar gauge singlets do not support symmetry non-restoration at arbitrary 
high temperatures. Does this phenomenon persists when considering gauged scalar fields? This  is the question we will answer in this 
section. We will find an example with the opposite behaviour, i.e. we will explicitly present 
a theory featuring two different gauge groups displaying simultaneously complete asymptotic freedom and symmetry non-restoration. 

To motivate the introduction of a second gauge group we will first show that with a single gauge group symmetries will restore at arbitrary high temperatures with(out) fermionic matter fields. 

Although the models in this section  share some features with the ones investigated in \cite{Chaudhuri:2020xxb}  the main difference resides in the fact that we are interested in symmetry non-restoration at arbitrary high temperatures. This means that we investigate theories near their UV fixed point, while in \cite{Chaudhuri:2020xxb} the authors concentrate on symmetry non-restoration occurring near interacting IR-fixed points.

\subsubsection{$SU(N_c)$ with $N_s$ fundamental scalars}

The $SU(N_c)\times SU(N_f)\times SU(N_s)$ symmetric Lagrangian is\footnote{If $N_c=N_s=4$ one can add to the potential 
a new invariant $w\,\det{S}+w^*\,\det{S^\dagger}$.}

\beq
{\cal L}=-\frac{1}{2}TrF_{\mu\nu}F^{\mu\nu}+Tr\left(\bar Qi\slashed{D}Q\right)+Tr\left(D^\mu SD_\mu S^\dagger \right)
-v\left(TrSS^\dagger\right)^2-u Tr\left(SS^\dagger\right)^2
\eeq

\noi
with the fields transforming as

\beq
Q\sim\left(N_c,N_f,1\right)\quad,\quad S\sim\left(N_c,1,N_s\right)
\eeq

The scalar thermal mass at high temperature is

\beq
\label{mS}
m_S^2(T)=(4\pi)^2\frac{T^2}{24\log{T}}\left(4(N_sN_c+1)\tilde\lambda_1+4(N_s+N_c)\tilde\lambda_2+3\frac{N_c^2-1}{N_c}\tilde\alpha\right)
\eeq

\noi
where we introduced, following \cite{Hansen:2017pwe},

\beq
v=(4\pi)^2\frac{\tilde\lambda_1}{t}\quad,\quad
u=(4\pi)^2\frac{\tilde\lambda_2}{t}\quad,\quad
g^2=(4\pi)^2\frac{\tilde\alpha}{t}
\eeq

\noi
with constant $\tilde\lambda_{1,2}$, $\tilde\alpha$.

The positivity of (\ref{mS}) follows from boundedness arguments. In fact  The $T=0$ potential is bounded from 
below iff \cite{Hansen:2017pwe}
\bea
\tilde\lambda_2\geq0&:&N_s\tilde\lambda_1+\tilde\lambda_2\geq0\\
\tilde\lambda_2\leq0&:&\tilde\lambda_1+\tilde\lambda_2\leq0
\eea
Since
\ben
\item
$\tilde\lambda_2\geq0$:
\beq
(N_sN_c+1)\tilde\lambda_1+(N_s+N_c)\tilde\lambda_2
=\frac{1}{N_s}\left(N_sN_c+1\right)\left(N_s\tilde\lambda_2+\tilde\lambda_2\right)
+\left(N_2-\frac{1}{N_s}\right)\tilde\lambda_2\geq0
\eeq
\item
$\tilde\lambda_2\geq0$:
\beq
(N_sN_c+1)\tilde\lambda_1+(N_s+N_c)\tilde\lambda_2
=(N_sN_c+1)\left(\tilde\lambda_1+\tilde\lambda_2\right)
+\left(N_s-1\right)\left(N_c-1\right)\left|\tilde\lambda_2\right|\geq0
\eeq
\een

\noi
we can now conclude that the thermal mass is always positive 

\beq
m_S^2(T)>0
\eeq

\noi
i.e. the symmetry is restored at high temperature.

\subsubsection{$SU(N_c)$ with two fundamental scalars}

One of the problems of the previous model was that there was too much symmetry in the scalar 
potential. We now take the case of two scalars, $N_s=2$, but instead of scalar SU(2) the symmetry 
of the potential will be just a discrete symmetry. We can take either a single $Z_2$ for even $N_c=2n$ 
or $Z_2\times Z_2$ for odd $N_c=2n+1$.

In fact:

\bet

\item
$Z_2\subset Z_{2n}$ and for even $N_c=2n$ the center of $SU(N_c)$ 
is $Z_{2n}$ and so $Z_2\subset SU(N_c)$. In other words, a common $\vec{\varphi}_i\to-\vec{\varphi}_i$ is 
already present. So in this case there is only one extra $Z_2$ possible, say $\vec{\varphi}_1\to-\vec{\varphi}_1$.

\item
for odd $N_c=2n+1$ there is no $Z_2$ subgroup of $SU(N_c)$. In fact using the Levi-Civita tensor 
the invariant out of $N_c=2n+1$ fundamentals is possible. Here it is thus possible to have an extra 
$Z_2\times Z_2$ symmetry for two fundamentals.

\eet

One way or another this means that each term of the potential can have only an even number of 
fundamentals $\vec{\varphi}_1$ and anti-fundamentals $\vec{\varphi}_1^*$ and an even number of 
fundamentals $\vec{\varphi}_2$ and anti-fundamentals $\vec{\varphi}_2^*$:

\bea
V&=&\frac{\lambda_1}{2}\left(\vec{\varphi}_1^*\cdot\vec{\varphi}_1\right)^2+\frac{\lambda_2}{2}\left(\vec{\varphi}_2^*\cdot\vec{\varphi}_2\right)^2+
\lambda_3\left(\vec{\varphi}_1^*\cdot\vec{\varphi}_1\right)\left(\vec{\varphi}_2^*\cdot\vec{\varphi}_2\right)\non\\
&+&\lambda_4\left(\vec{\varphi}_1^*\cdot\vec{\varphi}_2\right)\left(\vec{\varphi}_2^*\cdot\vec{\varphi}_1\right)+
\frac{\lambda_5}{2}\left(\vec{\varphi}_1^*\cdot\vec{\varphi}_2\right)^2+\frac{\lambda_5^*}{2}\left(\vec{\varphi}_2^*\cdot\vec{\varphi}_1\right)^2
\eea

\noi
with $\lambda_{1,2,3,4}$ real and in general $\lambda_5$ complex.

By taking 

\beq
g^2=\frac{16\pi^2\tilde\alpha}{N_ct}\quad,\quad \lambda_i=\frac{16\pi^2\tilde\lambda_i}{N_ct}
\eeq

\noi
with constant $\tilde\alpha$, $\tilde\lambda_i$, and

\beq
\tilde\lambda_\pm=\frac{1}{2}\left(\tilde\lambda_1\pm\tilde\lambda_2\right)
\eeq

\noi
the solutions to the RGE (see the appendix \ref{SUNc}) are

\ben
\item
\beq
\tilde\lambda_+=\frac{6\tilde\alpha-1}{4}\quad,\quad
\tilde\lambda_3^2+\tilde\lambda_-^2=\frac{24\tilde\alpha^2-12\tilde\alpha+1}{16}
\eeq
\item
\beq
\tilde\lambda_+=\frac{6\tilde\alpha-1+a_+\sqrt{24\tilde\alpha^2-12\tilde\alpha+1}}{4}\quad,\quad 
a_+^2=1\quad,\quad\tilde\lambda_3=\tilde\lambda_-=0
\eeq
\een

\noi
acceptable only for $\tilde\alpha\geq(3+\sqrt{3})/12$.

The thermal potential at large $N_c$ is
\beq
\Delta V_T=(4\pi)^2\frac{T^2}{24\log{T}}\left(\left(2\left(\tilde\lambda_1+\tilde\lambda_3\right)+3\tilde\alpha\right)
\left(\vec{\varphi}_1^*\cdot\vec{\varphi}_1\right)+\left(2\left(\tilde\lambda_2+\tilde\lambda_3\right)+3\tilde\alpha\right)
\left(\vec{\varphi}_2^*\cdot\vec{\varphi}_2\right)\right)
\eeq

The masses are quite symmetric and the search for symmetry restoration boils down to 
look for negative $\tilde\lambda_3=-|\tilde\lambda_3|$ which leads to a negative mass square for $\vec{\varphi}_1$, i.e. 
a negative combination

\beq
\label{function}
\frac{12\tilde\alpha-1}{2}-2\left(\sqrt{\frac{24\tilde\alpha^2-12\tilde\alpha+1}{16}-|\tilde\lambda_3|^2}+|\tilde\lambda_3|\right)
\eeq

\noi
for 

\beq
\tilde\alpha\geq\frac{3+\sqrt{3}}{12}\quad,\quad
0\leq|\tilde\lambda_3|\leq\sqrt{\frac{24\tilde\alpha^2-a2\tilde\alpha+1}{16}}
\eeq

The function (\ref{function}) is minimised for 

\beq
|\tilde\lambda_3|=\frac{1}{\sqrt{2}}\sqrt{\frac{24\tilde\alpha^2-12\tilde\alpha+1}{16}}
\eeq

\noi
which is however not enough for a negative mass square.

\subsubsection{$SU(N_{c_1}) \times SU(N_{c_2})$ with fundamental scalars: \\\ Symmetry breaks at high temperatures}

The model we will study now is similar to the previous one, but now we have two simple groups, $SU(N_{c_1}) \times SU(N_{c_2})$, 
so that each $\varphi_i$ is in a fundamentals 
representation of its $SU(N_{ci})$ and a singlet under the other one. 
The most general potential is

\beq
V=\frac{\lambda_1}{2}\left(\vec{\varphi}_1^*\cdot\vec{\varphi}_1\right)^2+\frac{\lambda_2}{2}\left(\vec{\varphi}_2^*\cdot\vec{\varphi}_2\right)^2-
\lambda\left(\vec{\varphi}_1^*\cdot\vec{\varphi}_1\right)\left(\vec{\varphi}_2^*\cdot\vec{\varphi}_2\right)
\eeq

Defining first

\bea
i=1,2&:&g_i^2=\frac{16\pi^2\tilde\alpha_i}{N_{ci}t}\quad,\quad \lambda_i=\frac{16\pi^2\tilde\lambda_i}{N_{ci}t}\\\
&&\lambda=\frac{16\pi^2\tilde\lambda}{\sqrt{N_{c1}N_{c2}}t}
\eea

\noi
with constant $\tilde\alpha_i$, $\tilde\lambda_i$, $\tilde\lambda$, the thermal effective potential becomes at large $N_{ci}$ reads

\begin{align}
\label{DVT}
\Delta V_T=
(4\pi)^2\frac{T^2}{24\log{T}}&\left(\left(2\left(\tilde\lambda_1-\sqrt{\frac{N_{c2}}{N_{c1}}}\tilde\lambda\right)+3\tilde\alpha_1\right)
\left(\vec{\varphi}_1^*\cdot\vec{\varphi}_1\right)\right.\non\\
&\left.+\left(2\left(\tilde\lambda_2-\sqrt{\frac{N_{c1}}{N_{c2}}}\tilde\lambda\right)+3\tilde\alpha_2\right)
\left(\vec{\varphi}_2^*\cdot\vec{\varphi}_2\right)\right)
\end{align}

Introducing the new variables

\beq
\tilde\lambda_\pm=\frac{1}{2}\left(\tilde\lambda_1\pm\tilde\lambda_2\right)\quad,\quad
\tilde\alpha_\pm=\frac{1}{2}\left(\tilde\alpha_1\pm\tilde\alpha_2\right)
\eeq

\noi
one finds the following solution\footnote{The other possible solution 
$\tilde\alpha_+=1/4$, $\tilde\lambda_+=\frac{1}{8}$, 
$\left(\tilde\lambda_--\frac{3}{2}\tilde\alpha_-\right)^2+\tilde\lambda^2=\frac{1}{32}\left(48\tilde\alpha_-^2-1\right)$ 
describes a $T=0$ potential which is unbounded from below.} 
of the RGE (see the appendix \ref{SUNc12}):

\bea
\tilde\alpha_-&=&0\\
\tilde\lambda_+&=&\frac{6\tilde\alpha_+-1}{4}\\
\tilde\lambda_-^2+\tilde\lambda^2&=&\frac{1}{16}\left(24\tilde\alpha_+^2-12\tilde\alpha_++1\right)
\eea

\noi
valid for 

\beq
\label{alphamin}
\tilde\alpha_+\geq(3+\sqrt{3})/12
\eeq

We will now prove that this solution supports symmetry non-restoration at arbitrary high temperatures. 

Denoting by $\mu^2_i$ the coefficient in front of $\left(\vec{\varphi}_i^*\cdot\vec{\varphi}_i\right)$ in 
the parenthesis on the right-hand-side of (\ref{DVT}) we have 

\bea
\mu_1^2&=&\frac{12\tilde\alpha_+-1}{2}+2\tilde\lambda_--2\sqrt{\frac{N_{c2}}{N_{c1}}}\tilde\lambda\\
\mu_2^2&=&\frac{12\tilde\alpha_+-1}{2}-2\tilde\lambda_--2\sqrt{\frac{N_{c1}}{N_{c2}}}\tilde\lambda
\eea
We are searching for positive 
\beq
\tilde\lambda=\left|\tilde\lambda\right|
\eeq

\noi
and, up to redefinitions of what is 1 and what is 2, we can take 

\beq
\tilde\lambda_-=-\sqrt{\frac{24\tilde\alpha_+^2-12\tilde\alpha_++1}{16}-\left|\tilde\lambda\right|^2}
\eeq

\noi
so that

\bea
\mu_1^2&=&\frac{12\tilde\alpha_+-1}{2}-2\left(\sqrt{\frac{24\tilde\alpha_+^2-12\tilde\alpha_++1}{16}-\left|\tilde\lambda\right|^2}
+\sqrt{\frac{N_{c2}}{N_{c1}}}\left|\tilde\lambda\right|\right)\\
\mu_2^2&=&\frac{12\tilde\alpha_+-1}{2}+2\left(\sqrt{\frac{24\tilde\alpha_+^2-12\tilde\alpha_++1}{16}-\left|\tilde\lambda\right|^2}
-\sqrt{\frac{N_{c1}}{N_{c2}}}\left|\tilde\lambda\right|\right)
\eea
Minimising the expression for $\mu_1^2$ we obtain: 
\beq
\left|\tilde\lambda\right|^2=\frac{1}{1+N_{c1}/N_{c2}}\frac{24\tilde\alpha_+^2-12\tilde\alpha_++1}{16}
\eeq

The minimised mass parameter

\beq
\mu_1^2=\frac{12\tilde\alpha_+-1}{2}-2\sqrt{\frac{24\tilde\alpha_+^2-12\tilde\alpha_++1}{16}}
\sqrt{1+\frac{N_{c2}}{N_{c1}}}
\eeq

\noi
can now be negative by a suitable choice of number of colours. 

\vskip .3cm
Let us now  demonstrate that the previous solution leads to a bounded potential. The latter occurs if 
\beq
\lambda_1\lambda_2-\lambda^2>0
\eeq

\noi
which  can be rewritten first as

\beq
\tilde\lambda_+^2-\tilde\lambda_-^2-\tilde\lambda^2>0
\eeq

\noi
and then as

\beq
\frac{(6\tilde\alpha_+-1)^2}{16}-\frac{24\tilde\alpha_+^2-12\tilde\alpha_++1}{16}=\frac{12\tilde\alpha_+^2}{16}
\eeq

\noi
which is indeed positive.  

Finally, requiring equal gauge couplings in the large $N_{ci}$ limit

\beq
\tilde\alpha_1=\tilde\alpha_2
\eeq

\noi
means that the original not rescaled couplings satisfy the relation

\beq
N_{c1}g_1^2=N_{c2}g_2^2
\eeq

This is achieved by the following suitable choice of number of matter fermions:

\beq
\frac{N_{f1}}{N_{c1}}=\frac{N_{f2}}{N_{c2}}
\eeq

Because of (\ref{alphamin}), they must satisfy

\beq
\frac{1}{2}\left(2+3\sqrt{3}\right)\leq\frac{N_{fi}}{N_{ci}}<\frac{11}{2}
\eeq

We arrive at the result, similar to Weinberg's model, that only one thermal mass is 
negative.

\subsubsection{The IR story of $SU(N_{c_1}) \times SU(N_{c_2})$ at nonzero temperature}

Interestingly the model of the previous subsection can feature also an IR Banks-Zaks fixed point. This can be achieved by tuning the number of fermions to maintain both gauge couplings equality and the occurrence o a perturbative IR 
fixed point. Once this is achieved the remaining equations for the IR fixed point values\footnote{N.B. These values should not to be confused with the 
tilded $1/t$ coefficients used for the fixed flow solutions in the UV.} are:
\bea
0&=&2\lambda_1^2+2\lambda^2-6\alpha_1\lambda_1+\frac{3}{2}\alpha_1^2\\
0&=&2\lambda_2^2+2\lambda^2-6\alpha_2\lambda_2+\frac{3}{2}\alpha_2^2\\
0&=&2\left(\lambda_1+\lambda_2\right)\lambda-3\left(\alpha_1+\alpha_2\right)\lambda
\eea

The solutions are 

\bea
\lambda_+&=&\frac{3}{2}\alpha_+\\
\lambda_-^2+\lambda^2&=&\frac{3}{2}\alpha_+^2
\eea

\noi
if and only if

\beq
\alpha_-=0
\eeq

The thermal effective potential is

\bea
\Delta V_T=
\frac{T^2}{24}&\times&\left(\left(2\left(\lambda_1-\sqrt{\frac{N_{c2}}{N_{c1}}}\lambda\right)+3\alpha_+\right)
\left(\vec{\varphi}_1^*\cdot\vec{\varphi}_1\right)\right.\non\\
&&\left.+\left(2\left(\lambda_2-\sqrt{\frac{N_{c1}}{N_{c2}}}\lambda\right)+3\alpha_+\right)
\left(\vec{\varphi}_2^*\cdot\vec{\varphi}_2\right)\right)
\eea

\noi
so that the thermal masses are proportional to

\bea
\mu_1^2&=&6\alpha_++2\lambda_--2\sqrt{\frac{N_{c2}}{N_{c1}}}\lambda\\
\mu_2^2&=&6\alpha_+-2\lambda_--2\sqrt{\frac{N_{c1}}{N_{c2}}}\lambda
\eea

Searching again for the branch

\beq
\lambda=\left|\lambda\right|
\quad,\quad\lambda_-=-\sqrt{\frac{3}{2}\alpha_+^2-\left|\lambda\right|^2}
\eeq

\noi
we have first

\bea
\mu_1^2&=&6\alpha_+-2\left(\sqrt{\frac{3}{2}\alpha_+^2-\left|\lambda\right|^2}
+\sqrt{\frac{N_{c2}}{N_{c1}}}\left|\lambda\right|\right)\\
\mu_2^2&=&6\alpha_++2\left(\sqrt{\frac{3}{2}\alpha_+^2-\left|\lambda\right|^2}
-\sqrt{\frac{N_{c1}}{N_{c2}}}\left|\lambda\right|\right)
\eea

$\mu_1^2$ is minimised for 

\beq
\left|\lambda\right|^2=\frac{1}{1+N_{c1}/N_{c2}}\frac{3}{2}\alpha_+^2
\eeq

\noi
so that the thermal mass

\beq
\mu_1^2=6\alpha_+-2\sqrt{\frac{3}{2}\alpha_+^2}
\sqrt{1+\frac{N_{c2}}{N_{c1}}}
\eeq

\noi
is negative for (but still in the Veneziano limit $N_{ci}\to\infty$)

\beq
\frac{N_{c2}}{N_{c1}}>5
\eeq 

Since

\beq
\lambda_+^2-\lambda_-^2-\lambda^2=\frac{3}{4}\alpha_+^2>0
\eeq

\noi
the parameter choice describes a $T=0$ potential which is bounded from below.

We have therefore found an example in which symmetry non-restoration occurs near an IR fixed point which is more minimal than the one presented in \cite{Chaudhuri:2020xxb}.

\subsubsection{Another example of symmetry breaking at high $T$: two adjoints in $SU(N_{c1})\times SU(N_{c2})$}

This model is similar to the previous one, except that adjoint scalars are considered instead of fundamental scalars. 
The details are described in appendix \ref{SUNc12adj}. The most general quartic potential is 

\beq
\label{V2adj}
V=\frac{\lambda_1^\prime}{4}Tr\Sigma_1^4+\frac{\lambda_2^\prime}{4}Tr\Sigma_2^4
+\frac{\lambda_1}{4}\left(Tr\Sigma_1^2\right)^2+\frac{\lambda_2}{4}\left(Tr\Sigma_2^2\right)^2
-\frac{\lambda}{2}Tr\Sigma_1^2Tr\Sigma_2^2
\eeq

We redefine the couplings as 

\bea
\lambda^\prime_{1,2}=(4\pi)^2\frac{\tilde\lambda^\prime_{1,2}}{N_{c_{1,2}}}\times\frac{1}{t}
&,&
\lambda_{1,2}=(4\pi)^2\frac{\tilde\lambda_{1,2}}{N_{c_{1,2}}^2}\times\frac{1}{t}\\
\lambda=(4\pi)^2\frac{\tilde\lambda}{N_{c_{1}}N_{c_{2}}}\times\frac{1}{t}
&,&g_{1,2}^2=(4\pi)^2\frac{\tilde\alpha_{1,2}}{N_{c_{1,2}}}\times\frac{1}{t}
\eea

\noi
with all tilded quantities constants, and eventually we will take the large $N_{c_{1,2}}$ limit.

As shown in \cite{Hansen:2017pwe}, the potential (\ref{V2adj}) is bounded from below 
if the parameters satisfy the following inequalities:

\beq
\lambda_i+\frac{\lambda_i^\prime}{k_i}>0
\quad\left(1\leq k_i\leq N_{c_i}\right)
\quad,\quad
\left(\lambda_1+\frac{\lambda_1^\prime}{k_1}\right)
\left(\lambda_2+\frac{\lambda_2^\prime}{k_2}\right)
>\lambda^2
\eeq

\noi
If $\lambda_i^\prime>0$, then it is enough to check the above for $k_i=N_{c_i}$, while if 
$\lambda_i^\prime<0$, $k_i=1$ suffices. However, for large $N_{c_i}$, the second case 
is impossible, since 

\beq
\lambda_i+\lambda_i^\prime>0\to\tilde\lambda_i^\prime>0
\eeq

\noi
which is in contradiction with the original assumption of $\lambda_i^\prime<0$.

\noi
So the only possibility is just $\lambda_1^\prime>0$, $\lambda_2^\prime>0$:

\beq
\tilde\lambda_1+\tilde\lambda_1^\prime>0
\quad,\quad
\tilde\lambda_2+\tilde\lambda_2^\prime>0
\quad,\quad
\left(\tilde\lambda_1+\tilde\lambda_1^\prime\right)
\left(\tilde\lambda_2+\tilde\lambda_2^\prime\right)
>\tilde\lambda^2
\eeq

The thermal mass is

\beq
V_T=(4\pi)^2\frac{T^2}{48\log{T}}\left(
\left(\tilde\lambda_1+2\tilde\lambda_1^\prime-
\frac{N_{c_2}}{N_{c_1}}\tilde\lambda+12\tilde\alpha_1\right)Tr\Sigma_1^2
+\left(\tilde\lambda_2+2\tilde\lambda_2^\prime-
\frac{N_{c_1}}{N_{c_2}}\tilde\lambda+12\tilde\alpha_2\right)Tr\Sigma_2^2\right)
\eeq

We provide here an existence proof for a negative thermal mass with parameters 
satisfying the boundedness of the potential constraint.

First one can show that only one sector would not work, as expected. This means that 
if $\tilde\lambda=\tilde\alpha_2=\tilde\lambda_2=\tilde\lambda_2^\prime=0$, there is no solution 
of the above fixed flow RG equations for real $\tilde\alpha_1$, $\tilde\lambda_1$, $\tilde\lambda_1^\prime$ assuming  
$\tilde\lambda_1+\tilde\lambda_1^\prime>0$ (boundedness) and 
$\tilde\lambda_1+2\tilde\lambda_1^\prime<0$ (negative thermal mass).

However, a solution for bounded potential with negative thermal mass square exists for 

\bea
\tilde\alpha_1=\tilde\alpha_2&=&\frac{2+\sqrt{2}}{2}\\
\tilde\lambda_1^\prime=\tilde\lambda_2^\prime&=&2\\
\tilde\lambda_1&=&12\left(2+\sqrt{2}\right)-26\\
\tilde\lambda_2&=&16\\
\tilde\lambda&=&\sqrt{120\left(2+\sqrt{2}\right)-392}\\
\frac{N_{c_2}}{N_{c_1}}&=&16 \ .
\eea

This is therefore another relevant example of symmetry non-restoration at arbitrary high temperature.

\section{Asymptotic safety at high temperature} 
Another way to achieve a UV complete theory, up to gravity, is via the presence of an interacting ultraviolet fixed point in all  couplings. In fact, one can have a combination of safe and free couplings for the model to be well defined at all scales. 

Due to the fact that the discovery of asymptotically safe quantum field theory is relatively recent \cite{Litim:2014uca} the issue of symmetry non-restoration for this relevant class of models has never been investigated before. 

We will consider here examples classified according to whether we can re-use part of the results  and reasoning  employed above for the complete asymptotically free theories or we need a separate in depth analysis of the safe model.  

For the first class  we consider theories structurally similar to the one considered above albeit with sufficient matter fields such that asymptotic freedom is lost while assuming  that perturbative asymptotic safety occurs. 

To transform the previous proof valid for asymptotically free theories to the equivalent  potential asymptotically safe case we need to  
\bet
\item
replace all tilded quantities with untilded ones;
\item
eliminate the $\log{T}$ in the denominator of the thermal mass;
\item
replace the $16\pi^2d\alpha_i/dt$ in the left-hand-sides of the RGEs with a zero.
\eet
This means that in the theories investigated in the previous section, once asymptotic freedom is lost and potential asymptotic safety appears, symmetry restoration is a must. 
 
\subsection{Explicit examples of asymptotic safety}
We now consider explicit constructions of asymptotically safe quantum field theories that cannot  be reduced to the example above because they either have multiple gauge singlet scalar quartic terms or/and have gauged scalars. Interestingly we anticipate that in both examples  the symmetry is restored at high temperature. 

 \subsubsection{The Litim-Sannino (LS) model}
The first model we consider here is the one put forward in \cite{Litim:2014uca} in which asymptotically safe quantum field theories and their structure was first discovered and understood. The Lagrangian reads: 

\bea
{\cal L}&=&-\frac{1}{2}Tr\left(F^{\mu\nu}F_{\mu\nu}\right)+Tr\left(\bar Qi\slashed{D} Q\right)
+Tr\left(\partial_\mu H^\dagger\partial^\mu H\right)\non\\
&+&y\,Tr\left(\bar Q_LHQ_R+\bar Q_RH^\dagger Q_L\right)-u\,Tr\left(H^\dagger H\right)^2-v\,\left(Tr H^\dagger H\right)^2 \ ,
\eea

\noi
with symmetry

\beq
G=SU(N_C)\times SU(N_F)\times SU(N_F) \times U_V(1) \ ,
\eeq

\noi
under which the fields transform as

\bea
Q_L&\sim&\left(N_C,N_F,1,1\right) \ ,\\
Q_R&\sim&\left(N_C,1,N_F,1\right) \ ,\\
H&\sim&\left(1,N_F,\overline{N_F},0\right) \ .
\eea

We assume the Veneziano limit, needed to ensure the rigorousness of the result 
\beq
N_F, N_C\to\infty \;\;\;,\;\;\;\frac{N_F}{N_C}=\frac{11}{2}+\epsilon \ ,
\eeq

\noi
with  $\epsilon \ll 1$ to control the size of the UV fixed point couplings that at the relevant order in perturbation theory read

\bea
\alpha_g&\equiv&\frac{g^2N_C}{(4\pi)^2}=\frac{26}{57}\epsilon+{\cal O}(\epsilon^2) \ ,\\
\alpha_y&\equiv&\frac{y^2N_C}{(4\pi)^2}=\frac{4}{19}\epsilon+{\cal O}(\epsilon^2)\ ,\\
\alpha_h&\equiv&\frac{uN_F}{(4\pi)^2}=\frac{\sqrt{23}-1}{19}\epsilon+{\cal O}(\epsilon^2)\ ,\\
\alpha_v&\equiv&\frac{vN_F^2}{(4\pi)^2}=-\frac{1}{19}\left(2\sqrt{23}-\sqrt{20+6\sqrt{23}}\right)\epsilon+{\cal O}(\epsilon^2) . 
\eea
The $T^2$ term of the $H$ mass square is
\bea
m_T^2&=&(4\pi)^2\,\frac{T^2}{24}\,\left(8\alpha_h+4\alpha_v+2\alpha_y\right)\non\\
&\approx&9.7\,\epsilon\,T^2>0 \ ,
\eea

\noi
so that the symmetry is restored at high temperature. Therefore we arrive at the conclusion that the original model of an asymptotically safe quantum field theory is also safe with respect to global symmetries. 

\subsection{A gauged scalar variant of the LS model}

Here we consider an interesting example featuring a two-scalar sector with one of the scalars being  gauged while the full theory  remains asymptotically safe   \cite{Abel:2018fls}. This model allows for a relevant test of symmetry (non)restoration and the Lagrangian of the model reads:  
\bea
{\cal L}&=&-\frac{1}{2}Tr\left(F^{\mu\nu}F_{\mu\nu}\right)+Tr\left(\bar Qi\slashed{D} Q\right)
+Tr\left(\partial_\mu H^\dagger\partial^\mu H\right)+Tr\left(D_\mu \tilde S^\dagger D^\mu \tilde S\right)\non\\
&+&\left(\frac{y}{\sqrt{2}}\,Tr\left(\tilde QHQ\right)+h.c.\right)-u_2\,Tr\left(H^\dagger H\right)^2-u_1\,\left(Tr H^\dagger H\right)^2\non\\
&-&w_2\,Tr\left(\tilde S^\dagger \tilde S\right)^2-w_1\,\left(Tr \tilde S^\dagger \tilde S\right)^2 \ ,
\eea

\noi
where the fields transform under the gauge and 3 global symmetries ($N_S=N_C-2$)

\beq
G=SU(N_C)\times SU(N_F)_L\times SU(N_F)_R\times SU(N_S) \ , 
\eeq

\noi
as

\bea
Q&\sim&(N_C,N_F,1,1) \ ,\\
\tilde Q&\sim&(\overline{N_C},1,\overline{N_F},1) \ ,\\
H&\sim&(1,\overline{N_F},N_F,1)\ ,\\
\tilde S&\sim&(\overline{N_C},1,1,\overline{N_S}) \ .
\eea

For small and positive 

\beq
\epsilon=\frac{N_F}{N_C}-\frac{11}{2}+\frac{N_S}{4N_C}\to\frac{N_F}{N_C}-\frac{21}{4} \ ,
\eeq

\noi
the following relations are satisfied \cite{Abel:2018fls} at the UV fixed point:

\bea
\alpha_g\equiv\frac{N_Cg^2}{(4\pi)^2}&=&\frac{25}{18}\epsilon \ ,\\
\alpha_y\equiv\frac{N_Cy^2}{(4\pi)^2}&=&\frac{24}{25}\alpha_g \ ,\\
\alpha_{u_1}\equiv\frac{N_F^2u_1}{(4\pi)^2}&=&\frac{-6\sqrt{22}+3\sqrt{19+6\sqrt{22}}}{100}\alpha_g \ ,\\
\alpha_{u_2}\equiv\frac{N_Fu_2}{(4\pi)^2}&=&\frac{3}{25}\left(\sqrt{22}-1\right)\alpha_g \ ,\\
\alpha_{w_1}\equiv\frac{N_C^2w_1}{(4\pi)^2}&=&\frac{3\pm\sqrt{3\left(4\sqrt{2}-5\right)}}{16\sqrt{2}}\alpha_g \ ,\\
\alpha_{w_2}\equiv\frac{N_Cw_2}{(4\pi)^2}&=&\frac{1}{16}\left(2-\sqrt{2}\right)\alpha_g \ .
\eea
Following the analysis of the LS case but now generalised to both scalars we arrive at

\bea
m_T^2(H)&=&(4\pi)^2\frac{T^2}{48}\left(2\alpha_y+16\alpha_{u_2}+8\alpha_{u_1}\right)\approx38.4\,\epsilon T^2>0 \ ,\\
m_T^2(S)&=&(4\pi)^2\frac{T^2}{24}\left(8\alpha_{w_2}+4\alpha_{w_1}+3\alpha_g\right)\approx37.2\,\epsilon T^2>0  \ .
\eea

\noi
This implies that no symmetries can be broken at high temperature. 

\section{Conclusions}

In this paper we analysed Weinberg's symmetry non-restoration idea within UV complete theories of  
either asymptotically free or safe nature.  

The reason why these are natural models to investigate is that  only for UV complete theories it is consistent to consider the arbitrary large temperature limit. 

Safe and free theories share short scale conformality that insures a well defined behaviour at arbitrary high energies. Because of this, they belong to a special subset of all possible quantum field theories. The remaining field theories should be considered as effective low energy descriptions that cannot be complete without quantum gravity possibly modifying their high energy behaviour. In any event, given the fact that we do not yet have a complete theory of quantum gravity, for these theories the symmetry non-restoration test cannot be performed at arbitrary high temperatures.  

As complete asymptotically free templates we commenced our investigation with $SU(N_c)$ gauge-Yukawa theories featuring $N_f$ fundamental Dirac fermions and two singlet scalars coupled via  Yukawa interactions to the fermions. We demonstrated that symmetry is restored for this class of asymptotically free theories. We then generalised the result to arbitrary 
(Weyl) fermion representations and to certain multiple singlet scalar theories. It was sufficient to demonstrate the incompatibility  between the request of negative thermal mass squared for one of the scalars and the simultaneous need for its coupling to be asymptotically free. 

We then moved to investigate the case of gauge scalars and have shown that high temperature symmetry non-restoration appeared for the case of two gauged scalars transforming according to the fundamental representation of two independent gauge sectors. Fermions in the fundamental representation were included as well but without Yukawa couplings.

We then moved to investigate the case of asymptotically safe theories starting by noticing that the  symmetry restoration results discovered for the singlet scalars  discussed above could be extended to potentially safe theories. 
 
Two more relevant examples were investigated in the asymptotically safe scenario in which either multiple quartic scalar field terms were present in the Lagrangian \cite{Litim:2014uca} and/or some of the scalar were gauged  \cite{Abel:2018fls}. In these models  symmetries restore at high temperature.  
 
 { 

As an interesting class of UV complete theories featuring symmetry non-restoration at arbitrary high temperatures we discovered the one featuring two gauged scalars, each in a 
fundamental representation of its own $SU(N_{ci})$ gauge group: for large enough ratios of colours, one scalar thermal mass can be 
negative.}

So far we discussed UV complete theories before adding quantum gravity.  We can imagine that a possible safe and free completion of the standard model occurs few orders of magnitude below the scale above which quantum gravity cannot be ignored. In this case our analysis still applies. It can even happen that quantum gravity is, per se, asymptotically free \cite{Salvio:2014soa}, and in this case we can ignore it.  

%

The simplicity of the UV complete models discovered here featuring arbitrary high temperature symmetry non-restoration phenomenon invites for further theoretical and phenomenological investigations. For example, it would be interesting to investigate whether UV complete grand-unified theories of the Pati-Salam type exist and that can feature the phenomenon of symmetry non-restoration.  Additionally there could be dark sectors that are gravitationally coupled to us that can be UV complete and feature early universe phase transitions from a symmetric to a broken one as the temperature increases. 
 
\subsubsection*{Acknowledgments}
BB is very grateful to Goran Senjanovi\'c for having been introduced to the fascinating subject of symmetry non-restoration at high temperature, as well as for illuminating interminable discussions  on related subjects. BB acknowledges the financial support from the Slovenian Research Agency 
(research core funding No.~P1-0035). 

\subsubsection*{Note added}
While we were completing the present work, a related paper appeared 
\cite{Chaudhuri:2020xxb} in which explicit examples of  Banks-Zaks type CFTs were considered in which symmetry nonrestoration occurred at nonzero temperature. 
Differently and in a complementary manner of \cite{Chaudhuri:2020xxb}  our work investigates, rather than theories around IR fixed points, models featuring either Gaussians (complete asymptotically free) or interacting  (completely asymptotically safe) UV  fixed points such that we can investigate the infinite temperature limit within a given UV complete quantum field theory.

\begin{appendices}

\section{\label{eqs} The 1-loop RG equations}
In this section we summarise the relevant one loop RG equations used in the main text starting with the normalisation of the fields  given by

\beq
{\cal L}_{kin}=-\frac{1}{4}F_{\mu\nu}^AF^{A\mu\nu}+i\overline{\Psi}\slashed{D}\Psi+\frac{1}{2}D^\mu\Phi^aD_\mu\Phi^a
\eeq

The gauge RG equation is

\beq
(4\pi)^2\beta_g\equiv(4\pi)^2\mu\frac{dg}{d\mu}=-b_0g^3
\eeq

\noi
with

\beq
b_0=\frac{11}{3}T(G)-\frac{2}{3}T(F)-\frac{1}{6}T(S)
\eeq

\noi
where $G,F,S$ stand for gauge bosons, Weyl fermions and real scalars, respectively, and $T(R)$ is the Dynkin 
index of the representation $R$, defined as 

\beq
Tr\left(T^A(R)T^B(R)\right)=T(R)\delta^{AB}
\eeq

In SU($N_c$) we will need the following:

\beq
T(fundamental)=\frac{1}{2}\;\;\;,\;\;\;T(adjoints)=N_c
\eeq

The Yukawa RG equations for Dirac fermions $\Psi_i$

\beq
\label{defYukawaDirac}
{\cal L}_{Yukawa}=\sum_{i,j}Y^a_{ij}\overline{\Psi}_i\phi^a\Psi_j
\eeq

\noi
are \cite{Machacek:1983fi} ($\kappa=1$ for Dirac fermions and $\kappa=1/2$ for Weyl fermions)

\bea
\label{machacekyukawa}
(4\pi)^2\beta_Y^a\equiv(4\pi)^2\mu\frac{dY^a}{d\mu}&=&\frac{1}{2}\left(Y^bY^{b\dagger}Y^a+Y^aY^{b\dagger}Y^b\right)+
2Y^bY^{a\dagger}Y^b\\
&+& \kappa Y^bTr\left(Y^{b\dagger}Y^a+Y^{a\dagger}Y^b\right)-3g^2\left(C_2(F)Y^a+Y^aC_2(F)\right)\non
\eea

\noi
where $\phi^a$ are real scalars and

\beq
\left(C_2(F)\right)_{ij}=\sum_{kA}T^A_{ik} T^A_{kj}
\eeq

\noi
where the generators $T^A$ are in the (in general reducible) representation of the fermions. 

Here and in the following a repeated index gets summed ($a$, $b$ over real scalars, $\alpha$ over SU($N_c$) generators, 
$i$, $j$, $k$ over (bi-)spinors) even when the explicit sum is not written.

Notice that the Yukawa matrices in (\ref{defYukawaDirac}) are Hermitean by definition.

The scalar sector is defined by

\beq
V=\frac{1}{4!}\lambda_{abcd}\phi_a\phi_b\phi_c\phi_d
\eeq

Following \cite{Machacek:1984zw} we introduce the completely symmetric tensors

\bea
\Lambda^2_{abcd}&=&\frac{1}{8}\sum_{perm}\lambda_{abef}\lambda_{efcd}\\
\Lambda^Y_{abcd}&=&\frac{1}{12}\sum_{perm}Tr\left(Y^{a\dagger}Y^e+Y^{e\dagger}Y^a\right)\lambda_{ebcd}\\
H_{abcd}&=&\frac{1}{4}\sum_{perm}Tr\left(Y^{a\dagger}Y^bY^{c\dagger}Y^d\right)\\
\Lambda^S_{abcd}&=&\frac{1}{6}\sum_{perm}\sum_{A=1}^{N_c^2-1}\left(T^A(S)T^A(S)\right)_{ae}\lambda_{ebcd}\\
A_{abcd}&=&\frac{1}{8}\sum_{perm}\sum_{A,B=1}^{N_c^2-1}\left\{T^A(S),T^B(S)\right\}_{ab}\left\{T^A(S),T^B(S)\right\}_{cd}
\eea

\noi
where the sum over "perm" means that we sum over all $4!$ permutations of the indices $a$, $b$, $c$ and $d$ 
so to make the left-hand sides completely symmetric in all indices. The matrices $T^A(S)$ are the Hermitean SU($N_c$) 
generators in the representation of the scalars. 
Since $\phi^a$ are taken real, these generators are imaginary and anti-symmetric. For real representations of SU($N_c$) 
this is automatic, while for complex representations one has to work out the form of these matrices. More precisely, they 
are found in the covariant derivative:

\beq
D_\mu\phi^a=\partial_\mu\phi^a-igW_{\mu}^A{\left(T^A(S)\right)^a}_b\phi^b
\eeq

For the case of more gauge couplings $g_\alpha$ of gauge groups with generators $T^A_{\alpha}$, one should remember that 

\beq
A_{abcd}\frac{\phi^a\phi^b\phi^c\phi^d}{4!}=\left(M^2_W\right)^{AB}\left(M^2_W\right)^{AB}
\eeq

\noi
with the $W$ mass

\beq
\left(M^2_W\right)^{AB}=\frac{1}{2}\phi^ag_{\alpha}g_{\beta}\left\{T_{\alpha}^A(S),T_{\beta}^B(S)\right\}_{ab}\phi^b
\eeq

The 1-loop RG equations then read \cite{Machacek:1984zw}

\beq
\label{machacekhiggs}
16\pi^2\frac{d\lambda_{abcd}}{dt}=\Lambda^2_{abcd}+2\kappa\Lambda^Y_{abcd}-8\kappa H_{abcd}
-3g^2\Lambda^S_{abcd}+3g^4A_{abcd}
\eeq

Finally, at high temperature the thermal mass matrix is given by  \cite{Weinberg:1974hy} (see also \cite{Dolan:1973qd})

\beq
\label{m2T}
m^2_{ab}(T)=\frac{T^2}{24}\left(\lambda_{abcc}+2\kappa Tr\left(Y^{a\dagger}Y^b+Y^{b\dagger}Y^a\right)+6g^2\left(T^A(S)T^A(S)\right)_{ab}\right)
\eeq

It is useful to rewrite the above formulae by multiplying the various quantities by constant $\phi^a\phi^b\phi^c\phi^d/4!$ and 
summing over the indices $a,b,c,d$. We thus define

\begin{align}
\label{VL2}
V_{\Lambda^2}&\equiv\Lambda^2_{abcd}\frac{\phi^a\phi^b\phi^c\phi^d}{4!}
=\frac{1}{2}\frac{\partial^2V}{\partial\phi^a\partial\phi^b}\frac{\partial^2V}{\partial\phi^a\partial\phi^b}\\
\label{VLY}
V_{\Lambda^Y}&\equiv2\kappa\Lambda^Y_{abcd}\frac{\phi^a\phi^b\phi^c\phi^d}{4!}
=\kappa\phi^a Tr\left(Y^{a\dagger}Y^e+Y^{e\dagger}Y^a\right)\frac{\partial V}{\partial\phi^e}\\
\label{VH}
V_H&\equiv-8\kappa H_{abcd}\frac{\phi^a\phi^b\phi^c\phi^d}{4!}
=-2\kappa Tr\left(Y^{a\dagger}Y^bY^{c\dagger}Y^d\right)\phi^a\phi^b\phi^c\phi^d\\
\label{VLS}
V_{\Lambda^S}&\equiv-3g^2\Lambda^S_{abcd}\frac{\phi^a\phi^b\phi^c\phi^d}{4!}
=-3g^2\phi^a\left(T^A(S)T^A(S)\right)_{ae}\frac{\partial V}{\partial\phi^e}\\
\label{VA}
V_A&\equiv3g^4A_{abcd}\frac{\phi^a\phi^b\phi^c\phi^d}{4!}
=\frac{3}{8}g^4\left(\phi^a\left\{T^A(S),T^B(S)\right\}_{ab}\phi^b\right)\left(\phi^c\left\{T^A(S),T^B(S)\right\}_{cd}\phi^d\right)
\end{align}

Eq. (\ref{machacekhiggs}) can thus be written as 

\beq
16\pi^2\frac{\phi^a\phi^b\phi^c\phi^d}{4!}\frac{d\lambda_{abcd}}{dt}=
V_{\Lambda^2}+V_{\Lambda^Y}+V_{H}+V_{\Lambda^S}+V_A
\eeq

\noi
while the equivalent of (\ref{m2T}) is (for vanishing Yukawa)

\beq
\Delta V(T)\equiv \frac{1}{2}m_{ab}^2(T)\phi^a\phi^b=\frac{T^2}{48}\left(2\frac{\partial^2V}{\partial\phi^a\partial\phi^a}+
6g^2\phi^a\left(T^A(S)T^A(S)\right)_{ab}\phi^b\right)
\eeq

\section{\label{SUNcsinglets}SU($N_c$) with two singlet scalars and fundamental fermions}
 
In this model the two singlet scalar couple through Yukawa couplings 
to $N_{f_1}$ ($N_{f_2}$) Dirac fermions in the fundamental representation of $SU(N_c)$. We further allow for $N_{f_0}$  
  Dirac fermions in the fundamental representation of the gauge group that are inert with respect to the scalars, i.e. do not possess Yukawa couplings. 

The gauge coupling 1-loop RGE is 

\beq
16\pi^2\frac{dg}{dt}=-b_0g^3\,,
\eeq

\noi
with

\beq
b_0=\frac{11}{3}N_c-\frac{2}{3}\left(N_{f_0}+N_{f_1}+N_{f_2}\right)\,.
\eeq

The solution is 

\beq
\alpha_g=\frac{g^2}{(4\pi)^2}=\frac{\tilde\alpha_g}{t}\,,
\eeq

\noi
with

\beq
\label{g}
\tilde\alpha_g=\frac{1}{2b_0}\,.
\eeq

The Yukawa RGE are

\beq
16\pi^2\frac{dy_i}{dt}=\left(3+2N_cN_{f_i}\right)y_i^3-3g^2\frac{N_c^2-1}{N_c}y_i\;\;\;,\;\;\;i=1,2\,.
\eeq

Assuming the ansatz

\beq
\alpha_{y_i}=\frac{y_i^2}{(4\pi)^2}=\frac{\tilde\alpha_{y_i}}{t}\,,
\eeq

\noi
the fixed flow solution is given by

\beq
\label{al12}
\tilde\alpha_{y_i}=\frac{6\frac{N_c^2-1}{N_c}\tilde\alpha_g-1}{2\left(3+2N_cN_{f_i}\right)}\;\;\;,\;\;\;i=1,2\,,
\eeq

\noi
and have positive solutions only if the gauge coupling is big enough

\beq
\label{pogojg}
6\tilde \alpha_g\frac{N_c^2-1}{N_c}-1>0\,,
\eeq

\noi
which reduces to a constraint on the number 
of Dirac fermion fundamentals :

\beq
\label{Nfinterval}
\frac{22}{4}N_c-\frac{9}{2}\left(N_c-\frac{1}{N_c}\right)<N_{f_0}+N_{f_1}+N_{f_2}<\frac{22}{4}N_c\,.
\eeq

The RG equations for the scalar couplings are

\bea
\label{d1}
16\pi^2\frac{d\lambda_1}{dt}&=&18\lambda_1^2+2\lambda^2
-8N_cN_{f_1}y_1^4+8N_cN_{f_1}y_1^2\lambda_1\\
16\pi^2\frac{d\lambda_2}{dt}&=&18\lambda_2^2+2\lambda^2
-8N_cN_{f_2}y_2^4+8N_cN_{f_2}y_2^2\lambda_2\\
\label{d}
16\pi^2\frac{d\lambda}{dt}&=&-8\lambda^2+6\lambda(\lambda_1+\lambda_2)
+4N_c\left(N_{f_1}y_1^2+N_{f_2}y_2^2\right)\lambda
\eea

The ansatz 

\beq
\alpha_{\lambda_i}=\frac{\lambda_i}{(4\pi)^2}=\frac{\tilde\alpha_{\lambda_i}}{t}\;\;\;,\;\;\;
\alpha_{\lambda}=\frac{\lambda}{(4\pi)^2}=\frac{\tilde\alpha_{\lambda}}{t}
\eeq

\noi
reduces the system of ODEs (\ref{d1})-(\ref{d}) to a system of algebraic equations

\bea
\label{eqlambda1}
-\tilde\alpha_{\lambda_1}&=&18\tilde\alpha_{\lambda_1}^2+2\tilde\alpha_{\lambda}^2
-8N_cN_{f_1}\tilde\alpha_{y_1}^2+8N_cN_{f_1}\tilde\alpha_{y_1}\tilde\alpha_{\lambda_1}\\
\label{eqlambda2}
-\tilde\alpha_{\lambda_2}&=&18\tilde\alpha_{\lambda_2}^2+2\tilde\alpha_{\lambda}^2
-8N_cN_{f_2}\tilde\alpha_{y_2}^2+8N_cN_{f_2}\tilde\alpha_{y_2}\tilde\alpha_{\lambda_2}\\
\label{eqlambda}
-\tilde\alpha_{\lambda}&=&-8\tilde\alpha_{\lambda}^2+6\tilde\alpha_{\lambda}\left(\tilde\alpha_{\lambda_1}
+\tilde\alpha_{\lambda_2}\right)+4N_c\left(N_{f_1}\tilde\alpha_{y_1}+N_{f_2}\tilde\alpha_{y_2}\right)\tilde\alpha_{\lambda}
\eea

To this we add (\ref{g}) and (\ref{al12}). We look for strictly positive solutions for all 6 couplings 
$\tilde\alpha_{g,y_1,y_2,\lambda_1,\lambda_2,\lambda}$, with 

\beq
\label{intervali}
N_c>1\;\;\;,\;\;\;N_{f_{0,1,2}}\geq0\;\;\;, \;\;
 N_{f_{1}}>0\;\; {\rm or}\;\; N_{f_2}>0 
\eeq

\noi
and $N_{f_0}+N_{f_1}+N_{f_2}$ in the interval (\ref{Nfinterval}).

Once this is obtained one can compute the thermal mass for the scalar scalars:

\beq
\label{mTNc}
m_i^2(T)=(4\pi)^2\frac{T^2}{12\log{T}}\left(3\tilde\alpha_{\lambda_i}-\tilde\alpha_{\lambda}+2N_cN_{f_i}\tilde\alpha_{y_i}\right)
\eeq

It turns out that there are 1784 inequivalent (we do not count those obtained by $N_{f_1}\leftrightarrow N_{f_2}$) 
choices of colours and flavours which satisfy (\ref{intervali}) and (\ref{Nfinterval}). However we are not only looking for 
fixed flow solutions, what we also need is that they lead to a negative thermal mass. 

We will now prove in general that there are no solutions with symmetry non-restoration. 

Let it be $m_1^2(T)<0$. To be so one needs

\beq
\label{pozitiven}
\tilde\alpha_{\lambda}-2N_cN_{f_1}\tilde\alpha_{y_1}>3\tilde\alpha_{\lambda_1}>0
\eeq 

We can now rewrite (\ref{eqlambda1}) as

\beq
\label{enacba}
2\left(\tilde\alpha_{\lambda}^2-4N_cN_{f_1}\tilde\alpha_{y_1}^2\right)+
\tilde\alpha_{\lambda_1}+18\tilde\alpha_{\lambda_1}^2
+8N_cN_{f_1}\tilde\alpha_{y_1}\tilde\alpha_{\lambda_1}=0
\eeq

All the terms except the first one are manifestly positive, so to satisfy the equation, the 
first term should be negative. However, the first term can be rewritten as

\beq
\tilde\alpha_{\lambda}^2-4N_cN_{f_1}\tilde\alpha_{y_1}^2=\left(\tilde\alpha_{\lambda}-2N_cN_{f_1}\tilde\alpha_{y_1}\right)
\left(\tilde\alpha_{\lambda}+2N_cN_{f_1}\tilde\alpha_{y_1}\right)+4N_cN_{f_1}\left(N_cN_{f_1}-1\right)\tilde\alpha_{y_1}^2
\eeq

This is positive, since the last term is non-negative, while the first product is positive due to (\ref{pozitiven}). Equation (\ref{enacba}) thus cannot 
have a solution. 

We  conclude the section summarising the result for the model presented:  {\it there is no fixed flow solution once a negative 
thermal mass is assumed}.

\section{Gauged scalars}

We consider in this appendix various examples of scalars in non-trivial representations of the gauge group.

\subsection{\label{TT} $SU(2)$ with two scalar triplets}

First we take the two scalar fields as gauge SU(2) triplets, coupled each to one fermion SU(2) doublet ($N_{f_1}=N_{f_2}=1$). 
To use almost all of the old results we still keep the $Z_2\times Z_2$ discrete symmetry. 
There is now an extra quartic term:

\beq
V=\frac{\lambda_1}{4}\left(\vec{\varphi}_1\cdot\vec{\varphi}_1\right)^2
+\frac{\lambda_2}{4}\left(\vec{\varphi}_2\cdot\vec{\varphi}_2\right)^2
-\frac{\lambda_{11}}{2}\left(\vec{\varphi}_1\cdot\vec{\varphi}_1\right)\left(\vec{\varphi}_2\cdot\vec{\varphi}_2\right)
-\frac{\lambda_{12}}{2}\left(\vec{\varphi}_1\cdot\vec{\varphi}_2\right)^2
\eeq

\noi
Denoting 

\beq
\phi=(\vec{\varphi_1},\vec{\varphi_2})
\eeq

\noi
we compute the quartic couplings directly from the definition

\beq
V=\frac{\lambda_{abcd}}{4!}\phi^a\phi^b\phi^c\phi^d
\eeq

\noi
i.e.

\beq
\lambda_{abcd}=\frac{\partial^4V}{\partial\phi^a\partial\phi^b\partial\phi^c\partial\phi^d}\;\;\;,\;\;\;a,b,c,d=1,\ldots,6
\eeq

For the Yukawa term we take

\beq
{\cal L}_{Yukawa}=\sum_{i=1}^2y_i\bar\psi_i\left(\frac{\vec{\tau}}{2}\cdot\vec{\varphi}_{i}\right)\psi_i
\eeq

\noi
with $\tau^A$, $A=1,2,3$ the Pauli matrices.

The (reducible) generators for the fermions (two fundamental representations of SU(2)) 

\beq
\Psi=(\psi_1,\psi_2)
\eeq

\noi
are

\beq
T^A=
\frac{1}{2}\left(
\begin{array}{cc}
 \tau^A & 0 \\
0 & \tau^A
\end{array}
\right)\;\;\;,\;\;\;A=1,2,3
\eeq

The fixed flow RGE are

\bea
\tilde\alpha_g&=&2b_0\tilde \alpha_g^2\\
\label{yukawa}
-\tilde\alpha_{y_i}&=&\frac{5}{2}\tilde\alpha_{y_i}^2-9\tilde\alpha_g\tilde\alpha_{y_i}\\
-\tilde\alpha_{\lambda _i}&=&
+22\tilde\alpha_{\lambda_i}^2
+6\tilde\alpha_{\lambda_{11}}^2
+4\tilde\alpha_{\lambda_{11}}\tilde\alpha_{\lambda_{12}}
+2\tilde\alpha_{\lambda_{12}}^2
-\tilde\alpha_{y_i}^2
+4\tilde\alpha_{\lambda_i}\tilde\alpha_{y_i}\non\\
&&+12 \tilde\alpha_g^2
-24\tilde\alpha_g\tilde\alpha_{\lambda_i}\\
 \tilde\alpha_{\lambda_{11}}&=&6\tilde\alpha_g^2+24\tilde\alpha_g\tilde\alpha_{\lambda_{11}}
+8\tilde\alpha_{\lambda_{11}}^2-10\tilde\alpha_{\lambda_{11}}(\tilde\alpha_{\lambda_1}+\tilde\alpha_{\lambda_2})\non\\
&&+2\tilde\alpha_{\lambda_{12}}^2-2\tilde\alpha_{\lambda_{12}}(\tilde\alpha_{\lambda_1}+\tilde\alpha_{\lambda_2})
-2\tilde\alpha_{\lambda_{11}}(\tilde\alpha_{y_1}+\tilde\alpha_{y_2})\\
\tilde\alpha_{\lambda_{12}}
&=&6\tilde\alpha_g^2+24\tilde\alpha_g\tilde\alpha_{\lambda_{12}}
+16\tilde\alpha_{\lambda_{11}}\tilde\alpha_{\lambda_{12}}\non\\
&&+10\tilde\alpha_{\lambda_{12}}^2
-4\tilde\alpha_{\lambda_{12}}(\tilde\alpha_{\lambda_1}+\tilde\alpha_{\lambda_2})
-2\tilde\alpha_{\lambda_{12}}(\tilde\alpha_{y_1}+\tilde\alpha_{y_2})
\eea

The thermal mass square results

\beq
m_i^2(T)=(4\pi)^2\frac{T^2}{12\log{T}}\left(\tilde\alpha_{y_i}+5\tilde\alpha_{\lambda_i}+6\tilde\alpha_g
-3\tilde\alpha_{\lambda_{11}}-\tilde\alpha_{\lambda_{12}}\right)
\eeq

The gauge beta function is known, 

\beq
b_0=\frac{22}{3}-\frac{2}{3}\left(N_{f_0}+2\right)-\frac{2}{3}=\frac{16-2N_{f_0}}{3}\to\tilde\alpha_g=\frac{3}{4(8-N_{f_0})}
\eeq

\noi
from where, to get $\tilde\alpha_g>1/9$, see (\ref{yukawa}), we need

\beq
2\leq N_{f_0}<8
\eeq

By explicit search on can find that there are no solutions of the fixed flow RGE for positive $\tilde\alpha_g$, $\tilde\alpha_{y_{1,2}}$, 
$\tilde\alpha_{\lambda_{1,2}}$ and real $\tilde\alpha_{\lambda_{11,12}}$.

\subsection{\label{ST} $SU(2)$ with one scalar singlet and one scalar triplet}

We take now one adjoint scalar and one singlet scalar that couple to the fermions (again in the fundamental representation, 
$N_{f_1}=N_{f_2}=1$) with the following Yukawa term: 

\beq
{\cal L}_{Yuk}=y_1\bar\psi_1{\phi}_{1} \psi_1 
+y_2\bar\psi_2\left(\frac{\vec{\tau}}{2}\cdot\vec{\varphi}_{2}\right)\psi_2 \ .
\eeq

Now the first scalar is singlet, the second is triplet. Obviously $\lambda_{12}$ cannot appear now. We will 
again call the remaining mixed constant $\lambda_{11}=\lambda$ in this section.

The fixed flow RGE are now

\bea
\tilde\alpha_g&=&2b_0\tilde \alpha_g^2\\
\label{yukawamixed1}
-\tilde\alpha_{y_1}&=&14\tilde\alpha_{y_1}^2-9\tilde\alpha_g \tilde\alpha_{y_1}\\
\label{yukawamixed2}
-\tilde\alpha_{y_2}&=&\frac{5}{2}\tilde\alpha_{y_2}^2-9\tilde\alpha_g\tilde\alpha_{y_2}\\
-\tilde\alpha_{\lambda_1}&=&18\tilde\alpha_{\lambda_1}^2+6\tilde\alpha_{\lambda}^2
-16\tilde\alpha_{y_1}^2+16\tilde\alpha_{\lambda_1}\tilde\alpha_{y_1}\\
-\tilde\alpha_{\lambda_2}&=&12\tilde\alpha_g^2-24\tilde\alpha_g\tilde\alpha_{\lambda_2}+2\tilde\alpha_{\lambda}^2
+22\tilde\alpha_{\lambda_2}^2-\tilde\alpha_{y_2}^2+4\tilde\alpha_{\lambda_2} \tilde\alpha_{y_2}\\
\tilde\alpha_{\lambda}&=&12\tilde\alpha_g\tilde\alpha_{\lambda}-6\tilde\alpha_{\lambda_1}\tilde\alpha_{\lambda}
+8\tilde\alpha_{\lambda}^2-10\tilde\alpha_{\lambda}\tilde\alpha_{\lambda_2}
-8\tilde\alpha_{\lambda}\left(\tilde\alpha_{y_1}+\frac{1}{4}\tilde\alpha_{y_2}\right)
\eea

\noi
while the thermal masses are

\bea
m_1^2(T)&=&(4\pi)^2\frac{T^2}{12\log{T}}\left(4\tilde\alpha_{y_1}+3\tilde\alpha_{\lambda_1}
-3\tilde\alpha_{\lambda}\right)\\
m_2^2(T)&=&(4\pi)^2\frac{T^2}{12\log{T}}\left(\tilde\alpha_{y_2}+5\tilde\alpha_{\lambda_2}+6\tilde\alpha_g
-\tilde\alpha_{\lambda}\right)
\eea

The gauge beta function is 

\beq
b_0=\frac{22}{3}-\frac{2}{3}\left(N_{f_0}+2\right)-\frac{1}{3}=\frac{17-2N_{f_0}}{3}\to\tilde\alpha_g=\frac{3}{2(17-2N_{f_0})}
\eeq

\noi
from where, to get $\tilde\alpha_g>1/9$, see (\ref{yukawamixed1}) or (\ref{yukawamixed2}), we need

\beq
2\leq N_{f_0}\leq8
\eeq

We find only two solutions:

\begin{align}
N_{f_0}=8&:
\left(\tilde\alpha_g,\tilde\alpha_{y_1},\tilde\alpha_{y_2},\tilde\alpha_{\lambda_1},\tilde\alpha_{\lambda_2},
\tilde\alpha_{\lambda}\right)=\left(1.5,0.893,5.0,0.518,0.182,0\right)\\
N_{f_0}=8&:
\left(\tilde\alpha_g,\tilde\alpha_{y_1},\tilde\alpha_{y_2},\tilde\alpha_{\lambda_1},\tilde\alpha_{\lambda_2},
\tilde\alpha_{\lambda}\right)=\left(1.5,0.893,5.0,0.518,0.5,0\right)
\end{align}

Since both have $\tilde\alpha_{\lambda}=0$, symmetry is always restored at high enough $T$.

\subsection{\label{SUNc}$SU(N_c)$ with two scalar fundamentals}

The potential is

\bea
V&=&\frac{\lambda_1}{2}\left(\vec{\varphi}_1^*\cdot\vec{\varphi}_1\right)^2+\frac{\lambda_2}{2}\left(\vec{\varphi}_2^*\cdot\vec{\varphi}_2\right)^2+
\lambda_3\left(\vec{\varphi}_1^*\cdot\vec{\varphi}_1\right)\left(\vec{\varphi}_2^*\cdot\vec{\varphi}_2\right)\non\\
&+&\lambda_4\left(\vec{\varphi}_1^*\cdot\vec{\varphi}_2\right)\left(\vec{\varphi}_2^*\cdot\vec{\varphi}_1\right)+
\frac{\lambda_5}{2}\left(\vec{\varphi}_1^*\cdot\vec{\varphi}_2\right)^2+\frac{\lambda_5^*}{2}\left(\vec{\varphi}_2^*\cdot\vec{\varphi}_1\right)^2
\eea

\noi
with $\lambda_{1,2,3,4}$ real and in general $\lambda_5$ complex.

The relation between the complex and the real basis is as usual

\beq
\varphi_{\alpha k}=\frac{1}{\sqrt{2}}\left(R_{\alpha k}+iI_{\alpha k}\right)\quad,\quad\alpha=1,2\quad,\quad k=1,\ldots,N_c
\eeq

\noi
so that 

\beq
\phi^a=\left(R_1^k,I_1^k,R_2^k,I_2^k\right)^T
\eeq

We get

\bea
V_{\Lambda^2}&=&
\frac{1}{2}\frac{\partial^2 V}{\partial\phi^a\partial\phi^b}\frac{\partial^2 V}{\partial\phi^b\partial\phi^a}\non\\
&=&\sum_{\alpha,\beta=1}^2\sum_{k,l=1}^{N_c}\left(
\frac{\partial^2 V}{\partial{\varphi_{\alpha}}^k\partial\varphi_{\beta l}^*}
\frac{\partial^2 V}{{\partial\varphi_{\beta}}^l\partial\varphi_{\alpha k}^*}+
\frac{\partial^2 V}{\partial{\varphi_{\alpha}}^k\partial{\varphi_{\beta}}^l}
\frac{\partial^2 V}{{\partial\varphi_{\beta l}}^*\partial\varphi_{\alpha k}^*}
\right)\\
&=&
Tr\left(M_1M_1^\dagger+2M_2M_2^\dagger+M_3M_3^\dagger+N_1N_1^\dagger+2N_2N_2^\dagger+N_3N_3^\dagger\right)\non
\eea

\noi
with

\bea
\label{M1}
M_1&=&\left(\lambda_1\left({\vec{\varphi}_1}^*\cdot\vec{\varphi}_1\right)
+\lambda_3\left(\vec{\varphi}_2^*\cdot\vec{\varphi}_2\right)\right)\mathbbm{1}
+\lambda_1\vec{\varphi}_1^*\otimes\vec{\varphi}_1
+\lambda_4\vec{\varphi}_2^*\otimes\vec{\varphi}_2\\
M_2&=&\left(\lambda_4\left(\vec{\varphi}_1^*\cdot\vec{\varphi}_2\right)
+\lambda_5^*\left(\vec{\varphi}_2^*\cdot\vec{\varphi}_1\right)\right)\mathbbm{1}
+\lambda_3\vec{\varphi}_1^*\otimes\vec{\varphi}_2
+\lambda_5^*\vec{\varphi}_2^*\otimes\vec{\varphi}_1\\
\label{M3}
M_3&=&\left(\lambda_2\left(\vec{\varphi}_2^*\cdot\vec{\varphi}_2\right)
+\lambda_3\left(\vec{\varphi}_1^*\cdot\vec{\varphi}_1\right)\right)\mathbbm{1}
+\lambda_4\vec{\varphi}_1^*\otimes\vec{\varphi}_1
+\lambda_2\vec{\varphi}_2^*\otimes\vec{\varphi}_2\\
N_1&=&\lambda_1\vec{\varphi}_1\otimes\vec{\varphi}_1
+\lambda_5\vec{\varphi}_2\otimes\vec{\varphi}_2\\
N_2&=&\lambda_3\vec{\varphi}_1\otimes\vec{\varphi}_2
+\lambda_4\vec{\varphi}_2\otimes\vec{\varphi}_1\\
N_3&=&\lambda_2\vec{\varphi}_2\otimes\vec{\varphi}_2
+\lambda_5^*\vec{\varphi}_1\otimes\vec{\varphi}_1
\eea

This gives

\begin{align}
V_{\Lambda^2}&=
\left(\left(2N_c+8\right)\lambda_1^2+2N_c\lambda_3^2+4\lambda_3\lambda_4+2\lambda_4^2+2\left|\lambda_5\right|^2\right)\frac{1}{2}
\left(\vec{\varphi}_1^*\cdot\vec{\varphi}_1\right)^2\non\\
&+\left(\left(2N_c+8\right)\lambda_2^2+2N_c\lambda_3^2+4\lambda_3\lambda_4+2\lambda_4^2+2\left|\lambda_5\right|^2\right)\frac{1}{2}
\left(\vec{\varphi}_2^*\cdot\vec{\varphi}_2\right)^2\non\\
&+\left(2\left(N_c+1\right)\left(\lambda_1+\lambda_2\right)\lambda_3+4\lambda_3^2+2\left(\lambda_1+\lambda_2\right)\lambda_4+2\lambda_4^2+2\left|\lambda_5\right|^2\right)
\left(\vec{\varphi}_1^*\cdot\vec{\varphi}_1\right)\left(\vec{\varphi}_2^*\cdot\vec{\varphi}_2\right)\non\\
&+\left(2\left(\lambda_1+\lambda_2\right)\lambda_4+8\lambda_3\lambda_4+2N_c\lambda_4^2+\left(4+2N_c\right)\left|\lambda_5\right|^2\right)
\left(\vec{\varphi}_1^*\cdot\vec{\varphi}_2\right)\left(\vec{\varphi}_2^*\cdot\vec{\varphi}_1\right)\non\\
&+2\left(\lambda_1+\lambda_2+4\lambda_3+2\left(N_c+1\right)\lambda_4\right)\lambda_5
\frac{1}{2}\left(\vec{\varphi}_1^*\cdot\vec{\varphi}_2\right)^2\non\\
&+2\left(\lambda_1+\lambda_2+4\lambda_3+2\left(N_c+1\right)\lambda_4\right)\lambda_5^*
\frac{1}{2}\left(\vec{\varphi}_2^*\cdot\vec{\varphi}_1\right)^2
\end{align}

We easily find

\bea
V_{\Lambda^S}&=&-3g^2\phi^a\left(T^A(S)T^A(S)\right)_{ab}\frac{\partial V}{\partial\phi^a}\non\\
&=&-3\frac{N_c^2-1}{2N_c}g^2
\left(\varphi_{\alpha}^k\frac{\partial V}{\partial\varphi_{\alpha}^k}+\varphi_{\alpha k}^{*}\frac{\partial V}{\partial\varphi_{\alpha k}^{*}}
\right)=-6\frac{N_c^2-1}{N_c}g^2V
\eea

Using

\beq
\phi^a\left\{T^A(S),T^B(S)\right\}_{ab}\phi^b=2g^2\varphi^*_{\alpha a}{\left\{T^A,T^B\right\}^a}_b\varphi_{\alpha}^b
\eeq

\noi
and the usual

\beq
{\left(T^A\right)^a}_b{\left(T^A\right)^c}_d=\frac{1}{2}\left({\delta^a}_d{\delta^c}_b-\frac{1}{N_c}{\delta^a}_b{\delta^c}_d\right)
\eeq

\noi
we get

\bea
V_A&=&\frac{3}{4}g^4\frac{N_c^2+2}{N_c^2}\left(\vec{\varphi}_1^*\cdot\vec{\varphi}_1+
\vec{\varphi}_2^*\cdot\vec{\varphi}_2\right)^2\non\\
&+&\frac{3}{4}g^4\frac{N_c^2-4}{N_c}\left(\left(\vec{\varphi}_1^*\cdot\vec{\varphi}_1\right)^2
+2\left(\vec{\varphi}_1^*\vec{\varphi}_2\right)\left(\vec{\varphi}_2^*\vec{\varphi}_1\right)+
\left(\vec{\varphi}_2^*\cdot\vec{\varphi}_2\right)^2\right)\non\\
&=&\frac{3}{4}g^4\frac{N_c^3+N_c^2-4N_c+2}{N_c^2}\left(\left(\vec{\varphi}_1^*\cdot\vec{\varphi}_1\right)^2+
\left(\vec{\varphi}_2^*\cdot\vec{\varphi}_2\right)^2\right)\\
&+&
\frac{3}{2}g^4\frac{N_c^2+2}{N_c^2}\left(\vec{\varphi}_1^*\cdot\vec{\varphi}_1\right)\left(\vec{\varphi}_2^*\cdot\vec{\varphi}_2\right)
+\frac{3}{2}g^4\frac{N_c^2-4}{N_c}\left(\vec{\varphi}_1^*\vec{\varphi}_2\right)\left(\vec{\varphi}_2^*\vec{\varphi}_1\right)\non
\eea

By taking 

\beq
g^2=\frac{16\pi^2\tilde\alpha}{N_ct}\quad,\quad \lambda_i=\frac{16\pi^2\tilde\lambda_i}{N_ct}
\eeq

\noi
we get for constant $\alpha,\lambda_i$ in the large $N_c$ limit the following fixed flow RGEs:

\bea
-\tilde\lambda_1&=&2\tilde\lambda_1^2+2\tilde\lambda_3^2-6\tilde\alpha\tilde\lambda_1+\frac{3}{2}\tilde\alpha^2\\
-\tilde\lambda_2&=&2\tilde\lambda_2^2+2\tilde\lambda_3^2-6\tilde\alpha\tilde\lambda_2+\frac{3}{2}\tilde\alpha^2\\
-\tilde\lambda_3&=&2\left(\tilde\lambda_1+\tilde\lambda_2\right)\tilde\lambda_3-6\tilde\alpha\tilde\lambda_3\\
-\tilde\lambda_4&=&2\tilde\lambda_4^2+2\left|\tilde\lambda_5\right|^2-6\tilde\alpha\tilde\lambda_4+\frac{3}{2}\tilde\alpha^2\\
-\tilde\lambda_5&=&4\tilde\lambda_4\tilde\lambda_5-6\tilde\alpha\tilde\lambda_5
\eea

The thermal potential is

\bea
\Delta V_T&=&
\frac{T^2}{48}\left(2\frac{\partial^2V}{\partial\phi^a\partial\phi^a}+6\phi^a\left(T^A(S)T^A(S)\right)_{ab}\phi^b\right)\non\\
&=&\frac{T^2}{48}\sum_{i=1}^2\sum_{a=1}^{N_c}\left(4\frac{\partial^2V}{\partial\varphi_i^a\partial\varphi_{ia}^*}
+6g^2\frac{N_c^2-1}{N_c}\varphi_i^a\varphi_{ia}^*\right)\non\\
\eea

Using (\ref{M1}) and (\ref{M3}) 

\bea
\sum_{i=1}^2\sum_{a=1}^{N_c}\frac{\partial^2V}{\partial\varphi_i^a\partial\varphi_{ia}^*}&=&
Tr\left(M_1+M_3\right)\non\\
&=&\left(\left(N_c+1\right)\lambda_1+N_c\lambda_3+\lambda_4\right)\left(\vec{\varphi}_1^*\cdot\vec{\varphi}_1\right)\non\\
&+&\left(\left(N_c+1\right)\lambda_2+N_c\lambda_3+\lambda_4\right)\left(\vec{\varphi}_2^*\cdot\vec{\varphi}_2\right)
\eea

At large $N_c$ 

\beq
\Delta V_T=(4\pi)^2\frac{T^2}{24\log{T}}\left(\left(2\left(\tilde\lambda_1+\tilde\lambda_3\right)+3\tilde\alpha\right)
\left(\vec{\varphi}_1^*\cdot\vec{\varphi}_1\right)+\left(2\left(\tilde\lambda_2+\tilde\lambda_3\right)+3\tilde\alpha\right)
\left(\vec{\varphi}_2^*\cdot\vec{\varphi}_2\right)\right)
\eeq

\subsection{\label{SUNc12} $SU(N_{c_1}) \times SU(N_{c_2})$  with two scalar fundamentals}

The model we will study now is similar to the previous one, but now we have two simple groups, $SU(N_{c_1}) \times SU(N_{c_2})$, 
so that each $\varphi_i$ is in a fundamentals 
representation of its $SU(N_{ci})$ and a singlet under the other one. 
The most general potential is

\beq
V=\frac{\lambda_1}{2}\left(\vec{\varphi}_1^*\cdot\vec{\varphi}_1\right)^2+\frac{\lambda_2}{2}\left(\vec{\varphi}_2^*\cdot\vec{\varphi}_2\right)^2-
\lambda\left(\vec{\varphi}_1^*\cdot\vec{\varphi}_1\right)\left(\vec{\varphi}_2^*\cdot\vec{\varphi}_2\right)
\eeq

As before we derive the various pieces of the RGE using (\ref{VL2}), (\ref{VLS}), (\ref{VA}):

\bea
V_{\Lambda^2}&=&
\left(\left(2N_{c1}+8\right)\lambda_1^2+2N_{c2}\lambda^2\right)\frac{1}{2}
\left(\vec{\varphi}_1^*\cdot\vec{\varphi}_1\right)^2\non\\
&&+\left(\left(2N_{c2}+8\right)\lambda_2^2+2N_{c1}\lambda^2\right)\frac{1}{2}
\left(\vec{\varphi}_2^*\cdot\vec{\varphi}_2\right)^2\\
&&-\left(2\left(N_{c1}\lambda_1+N_{c2}\lambda_2\right)\lambda+2\left(\lambda_1+\lambda_2\right)\lambda-4\lambda^2\right)
\left(\vec{\varphi}_1^*\cdot\vec{\varphi}_1\right)\left(\vec{\varphi}_2^*\cdot\vec{\varphi}_2\right)\non\\
V_{\Lambda^S}&=&
-6\frac{N_{c1}^2-1}{N_{c1}}g_1^2\left(\frac{\lambda_1}{2}\left(\vec{\varphi}_1^*\cdot\vec{\varphi}_1\right)^2
-\frac{\lambda}{2}\left(\vec{\varphi}_1^*\cdot\vec{\varphi}_1\right)\left(\vec{\varphi}_2^*\cdot\vec{\varphi}_2\right)\right)\non\\
&&
-6\frac{N_{c2}^2-1}{N_{c2}}g_2^2\left(\frac{\lambda_2}{2}\left(\vec{\varphi}_2^*\cdot\vec{\varphi}_2\right)^2
-\frac{\lambda}{2}\left(\vec{\varphi}_1^*\cdot\vec{\varphi}_1\right)\left(\vec{\varphi}_2^*\cdot\vec{\varphi}_2\right)\right)\\
V_{A}&=&\frac{3}{4}g_1^4\frac{N_{c1}^3+N_{c1}^2-4N_{c1}+2}{N_{c1}^2}
\left(\vec{\varphi}_1^*\cdot\vec{\varphi}_1\right)^2\non\\
&&+\frac{3}{4}g_2^4\frac{N_{c2}^3+N_{c2}^2-4N_{c2}+2}{N_{c2}^2}
\left(\vec{\varphi}_2^*\cdot\vec{\varphi}_2\right)^2
\eea

Defining

\bea
i=1,2&:&g_i^2=\frac{16\pi^2\tilde\alpha_i}{N_{ci}t}\quad,\quad \lambda_i=\frac{16\pi^2\tilde\lambda_i}{N_{ci}t}\\\
&&\lambda=\frac{16\pi^2\tilde\lambda}{\sqrt{N_{c1}N_{c2}}t}
\eea

\noi
with constant 
we get for the RG equations at large $N_{ci}$

\bea
-\tilde\lambda_1&=&2\tilde\lambda_1^2+2\tilde\lambda^2-6\tilde\alpha_1\tilde\lambda_1+\frac{3}{2}\tilde\alpha_1^2\\
-\tilde\lambda_2&=&2\tilde\lambda_2^2+2\tilde\lambda^2-6\tilde\alpha_2\tilde\lambda_2+\frac{3}{2}\tilde\alpha_2^2\\
-\tilde\lambda&=&2\left(\tilde\lambda_1+\tilde\lambda_2\right)\tilde\lambda-3\left(\tilde\alpha_1+\tilde\alpha_2\right)\tilde\lambda
\eea

The thermal effective potential 

\beq
\Delta V_T=
\frac{T^2}{48}\sum_{i=1}^2\sum_{a=1}^{N_c}\left(4\frac{\partial^2V}{\partial\varphi_i^a\partial\varphi_{ia}^*}
+6g_i^2\frac{N_{ci}^2-1}{N_{ci}}\varphi_i^a\varphi_{ia}^*\right)\non\\
\eeq

\noi
becomes at large $N_{ci}$ 

\begin{align}
\Delta V_T=
(4\pi)^2\frac{T^2}{24\log{T}}&\left(\left(2\left(\tilde\lambda_1-\sqrt{\frac{N_{c2}}{N_{c1}}}\tilde\lambda\right)
+3\tilde\alpha_1\right)
\left(\vec{\varphi}_1^*\cdot\vec{\varphi}_1\right)\right.\non\\
&\left.+\left(2\left(\tilde\lambda_2-\sqrt{\frac{N_{c1}}{N_{c2}}}\tilde\lambda\right)+3\tilde\alpha_2\right)
\left(\vec{\varphi}_2^*\cdot\vec{\varphi}_2\right)\right)
\end{align}

\subsection{\label{SUNc12adj} $SU(N_{c_1}) \times SU(N_{c_2})$  with two scalar adjoints}

We now present a model again with two simple gauge groups, $SU(N_{c_1}) \times SU(N_{c_2})$, and one adjoint for 
each gauge group. The potential is parametrised by 

\beq
V=\frac{\lambda_1^\prime}{4}Tr\Sigma_1^4+\frac{\lambda_2^\prime}{4}Tr\Sigma_2^4
+\frac{\lambda_1}{4}\left(Tr\Sigma_1^2\right)^2+\frac{\lambda_2}{4}\left(Tr\Sigma_2^2\right)^2
-\frac{\lambda}{2}Tr\Sigma_1^2Tr\Sigma_2^2
\eeq

The 1-loop corrections are (\ref{VL2}), (\ref{VLS}), (\ref{VA}):

\begin{align}
V_{\Lambda^2}&=\frac{1}{8}\left(
Tr\Sigma_1^4\left(12\lambda_1\lambda_1^\prime+\lambda_1^{\prime 2}\frac{2N_{c_1}^2-18}{N_{c_1}}\right)
-Tr\Sigma_1^2Tr\Sigma_2^2\left(\lambda_1\lambda\left(2N_{c_1}^2+
2\right)+\lambda\lambda_1^\prime\frac{4N_{c_1}^2-6}{N_{c_1}}\right)\right.\non\\
&\left.+\left(Tr\Sigma_1^2\right)^2\left(\lambda_1^2\left(N_{c_1}^2+7\right)+\lambda_1\lambda_1^\prime\frac{4N_{c_1}^2-6}{N_{c_1}}
+\lambda_1^{\prime 2}\frac{3N_{c_1}^2+9}{N_{c_1}^2}\right)+\left(Tr\Sigma_2^2\right)^2\lambda^2\left(N_{c_1}^2-1\right)\right)\non\\
&+Tr\Sigma_1^2Tr\Sigma_2^2\lambda^2\\
&+\frac{1}{8}\left(
Tr\Sigma_2^4\left(12\lambda_2\lambda_2^\prime+\lambda_2^{\prime 2}\frac{2N_{c_2}^2-18}{N_{c_2}}\right)
-Tr\Sigma_1^2Tr\Sigma_2^2\left(\lambda_2\lambda\left(2N_{c_2}^2+
2\right)+\lambda\lambda_2^\prime\frac{4N_{c_2}^2-6}{N_{c_2}}\right)\right.\non\\
&\left.+\left(Tr\Sigma_2^2\right)^2\left(\lambda_2^2\left(N_{c_2}^2+7\right)+\lambda_2\lambda_2^\prime\frac{4N_{c_2}^2-6}{N_{c_2}}
+\lambda_2^{\prime 2}\frac{3N_{c_2}^2+9}{N_{c_2}^2}\right)+\left(Tr\Sigma_1^2\right)^2\lambda^2\left(N_{c_2}^2-1\right)\right)\non
\end{align}

\begin{align}
V_{\Lambda^S}=-&3g_1^2N_{c_1}\left(\lambda_1^\prime Tr\Sigma_1^4+\lambda_1\left(Tr\Sigma_1^2\right)^2-
\lambda Tr\Sigma_1^2Tr\Sigma_2^2\right)\non\\
-&3g_2^2N_{c_2}\left(\lambda_2^\prime Tr\Sigma_2^4+\lambda_2\left(Tr\Sigma_2^2\right)^2-
\lambda Tr\Sigma_1^2Tr\Sigma_2^2\right)
\end{align}

\beq
V_{\Lambda_A}=3g_1^4
\left(N_{c_1}Tr\Sigma_1^4+3\left(Tr\Sigma_1^2\right)^2\right)+3g_2^4
\left(N_{c_2}Tr\Sigma_2^4+3\left(Tr\Sigma_2^2\right)^2\right)
\eeq

We redefine the constants as 

\bea
\lambda^\prime_{1,2}=(4\pi)^2\frac{\tilde\lambda^\prime_{1,2}}{N_{c_{1,2}}}\times\frac{1}{t}
&,&
\lambda_{1,2}=(4\pi)^2\frac{\tilde\lambda_{1,2}}{N_{c_{1,2}}^2}\times\frac{1}{t}\\
\lambda=(4\pi)^2\frac{\tilde\lambda}{N_{c_{1}}N_{c_{2}}}\times\frac{1}{t}
&,&g_{1,2}^2=(4\pi)^2\frac{\tilde\alpha_{1,2}}{N_{c_{1,2}}}\times\frac{1}{t}
\eea

\noi
with all tilded quantities constants, and eventually took the large $N_{c_{1,2}}$ limit.

In the Veneziano limit the RGE are

\bea
-\tilde\lambda_1&=&\frac{1}{2}\tilde\lambda_1^2+2\tilde\lambda_1\tilde\lambda_1^\prime
+\frac{3}{2}\tilde\lambda_1^{\prime 2}+\frac{1}{2}\tilde\lambda^2-12\tilde\alpha_1\tilde\lambda_1+36\tilde\alpha_1^2\\
-\tilde\lambda_2&=&\frac{1}{2}\tilde\lambda_2^2+2\tilde\lambda_2\tilde\lambda_2^\prime
+\frac{3}{2}\tilde\lambda_2^{\prime 2}+\frac{1}{2}\tilde\lambda^2-12\tilde\alpha_2\tilde\lambda_2+36\tilde\alpha_2^2\\
-\tilde\lambda&=&\tilde\lambda\left(\frac{1}{2}\left(\tilde\lambda_1+\tilde\lambda_2\right)+\tilde\lambda_1^{\prime}+\tilde\lambda_2^{\prime}-6\left(\tilde\alpha_1+\tilde\alpha_2\right)\right)\\
-\tilde\lambda_1^\prime&=&\tilde\lambda_1^{\prime 2}-12\tilde\alpha_1\tilde\lambda_1^\prime+12\tilde\alpha_1^2\\
-\tilde\lambda_2^\prime&=&\tilde\lambda_2^{\prime 2}-12\tilde\alpha_2\tilde\lambda_2^\prime+12\tilde\alpha_2^2
\eea

The thermal mass is

\begin{align}
V_T=\frac{T^2}{48}&\left(
\left(\lambda_1\left(N_{c_1}^2+1\right)+\lambda_1^\prime\frac{2N_{c_1}^2-3}{N_{c_1}}-
\lambda\left(N_{c_2}^2-1\right)+12N_{c_1}g_1^2\right)Tr\Sigma_1^2\right.\non\\
&\left.
\left(\lambda_2\left(N_{c_2}^2+1\right)+\lambda_2^\prime\frac{2N_{c_2}^2-3}{N_{c_2}}-
\lambda\left(N_{c_1}^2-1\right)+12N_{c_2}g_2^2\right)Tr\Sigma_2^2\right)
\end{align}

\noi
and becomes in the Veneziano limit

\beq
V_T=(4\pi)^2\frac{T^2}{48\log{T}}\left(
\left(\tilde\lambda_1+2\tilde\lambda_1^\prime-
\frac{N_{c_2}}{N_{c_1}}\tilde\lambda+12\tilde\alpha_1\right)Tr\Sigma_1^2
+\left(\tilde\lambda_2+2\tilde\lambda_2^\prime-
\frac{N_{c_1}}{N_{c_2}}\tilde\lambda+12\tilde\alpha_2\right)Tr\Sigma_2^2\right)
\eeq

\end{appendices}

\end{document}